\documentclass{psp-rv975x65}
\usepackage{psp-rv-van}             
\usepackage{subfigure}
\usepackage[dvips]{color}

\newcommand{\bb}{\mbox{{\bf b}}}
\newcommand{\fb}{\mbox{{\bf f}}}
\newcommand{\Mb}{\mbox{{\bf M}}}
\newcommand{\ub}{\mbox{{\bf u}}}
\newcommand{\nhat}{\mbox{\boldmath{$\hat{n}$}}}
\newcommand{\sigmab}{\mbox{{\boldmath{$\sigma$}}}}

\def\p{\par $\bullet$ }

\makeindex
\copyline{Nanoscale liquid interfaces}{2010}{978-981-nnnn-nn-n}
\begin{document}

\chapter[Wetting, roughness and hydrodynamic slip]{Wetting, roughness and hydrodynamic slip\label{ch1}}

\author{Olga I. Vinogradova$^{1,2,3}$ and Aleksey V. Belyaev$^{1,2}$}

\address{$^1$ A.N.~Frumkin Institute of Physical Chemistry and Electrochemistry, Russian Academy of Sciences, 31 Leninsky Prospect, 119991 Moscow, Russia \\
$^2$ Physics Department, M.V.~Lomonosov Moscow State University, 119991 Moscow, Russia \\
$^3$ ITMC und DWI an der RWTH Aachen, Pauwelsstr. 8,
52056 Aachen, Germany
}


\begin{abstract}
 The hydrodynamic slippage at a solid-liquid
interface is currently at the center of our understanding
of fluid mechanics. For hundreds of years this science has relied upon no-slip boundary conditions at the solid-liquid interface that has been applied successfully to model many
macroscopic experiments, and the  state of this interface has played a minor role in determining the flow. However, the problem is not that simple and has been revisited recently. Due to the change in the properties of the interface, such as wettability and roughness, this classical boundary condition could be violated, leading to a hydrodynamic slip. In this chapter, we review recent advances in the understanding and expectations for the hydrodynamic boundary conditions in different situations, by focussing mostly on key papers from past decade. We highlight mostly the impact of hydrophobicity, roughness, and especially their combination on the flow properties. In particular, we show that hydrophobic slippage can be dramatically affected by the presence of roughness, by inducing
novel hydrodynamic phenomena, such as giant interfacial slip, superfluidity,
mixing, and low hydrodynamic drag. Promising directions for further research are also discussed.
\end{abstract}


\body

\section{Introduction}

\begin{figure}
\begin{center}
  \includegraphics [width=0.8\textwidth]{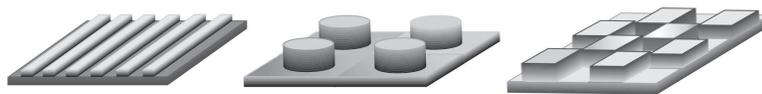}
  \end{center}
  \caption{Examples of anisotropic (grooves) and isotropic (pillars, chess-board) textures.}
  \label{fig:texture}
\end{figure}

Fluid mechanics is one of the oldest and useful of the `exact' sciences. For hundreds of years it has relied upon the no-slip boundary condition at a solid-liquid interface, that was applied successfully to model many
macroscopic experiments~\cite{batchelor.gk:2000}. However, the problem is not that simple and has been revisited during recent years. One reason for such a strong interest in `old' problem is purely fundamental. The no-slip boundary condition is an assumption that cannot be derived from first principles even for a molecularly smooth hydrophilic [the contact angle (fixed by the chemical nature of a solid) lies between 0$^\circ$ and 90$^\circ$] surface. Therefore, the success of no-slip postulate may not
always reflect its accuracy but in fact rather the insensitivity of the experiment. Another reason for current interest to flow boundary conditions lies in the potential applications in many areas of engineering and applied science, which deal with small size systems, including micro- and nanofluidics~\cite{squires2005}, flow in porous media, friction and lubrication, and biological fluids. The driving and mixing of liquids when the channel size decreases represent a very difficult problem~\cite{stone2004}. There is therefore a big hope to cause changes in hydrodynamic behavior by an impact of interfacial phenomena on the flow. For example, even ideal solids, which are both flat and chemically homogeneous, can have a contact angle, which exceeds 90$^\circ$ (the hydrophobic case). This can modify hydrodynamic boundary conditions, as it has been shown yet in early work~\cite{vinogradova1999}. Besides that, solids are not ideal, yet rough. This can further change, and quite dramatically, boundary conditions.
It is of course interesting and useful to show how the defects or pores of the solids modify
them. But today, the question has slightly shifted. Thanks to techniques coming
from microelectronics, we are able to elaborate substrates whose surfaces are patterned (often
at the micro- and nanometer scale) in a very well controlled way (see Fig.~\ref{fig:texture}), which provides properties (e.g. optical
or electrical) that the solid did not have when flat or slightly disordered. A texture affects the wettability and boundary conditions on a substrate, and can induce unique properties
that the material could not have without these micro- and nanostructures. In particular, in case of super-hydrophobic solids, which are generated by a combination of surface chemistry
and patterns, roughness can dramatically lower the ability of drops to stick, by leading to the remarkable mobility of liquids. At the macroscopic scale this renders them `self-cleaning' and causes droplets to roll (rather
than slide) under gravity and rebound (rather than spread) upon
impact instead of spreading~\cite{quere.d:2005}. At the smaller scale, reduced wall friction and a superlubricating potential are almost likely associated with the breakdown of the no-slip
hypothesis. This suggests that super-hydrophobic surfaces should strongly affect the transport of fluids. This area of research is relatively new, but rapidly developing, and it already attracted scientists from physics, chemistry, mathematics, and engineering.

In this chapter we review recent advances in the understanding and expectations for the fluid-solid boundary conditions in different situations, where hydrophobicity and roughness impact the flow properties, by focussing mostly on key papers from past decade. Throughout the chapter, we emphasize open questions. We start with a very brief history of the no-slip boundary conditions. Then, after introducing the terminology and models, and describing new developments and instruments, which give the possibility of investigating fluid behavior at the micro- and nanoscale, in the following section we present results obtained for smooth surfaces, by highlighting the role of wettability. Then follows the results for a rough hydrophilic  and, especially,  hydrophobic surfaces. In the latter case we show how can roughness enhance hydrodynamic slip and thus the efficiency of interfacial transport phenomena.  We close with the suggestions to guide the next round of experimental and theoretical studies in this expanding area of research.

\section{Origins and history}

The nature of boundary conditions in hydrodynamics was widely debated in 19th century, and many of the great names in fluid dynamics have expressed an opinion on the subject, as discussed in a recent review.~\cite{lauga2005} The linear slip boundary condition was introduced by Navier,~\cite{navier1823} and this remains a standard characteristic of slip used today. Helmholtz and von Piotrowski were probably the first to report some evidence
of slippage at the solid-liquid interface~\cite{helmholtz.h:1860}. We refer the reader to a
comprehensive review~\cite{binhgam.ec:1922} for detailed accounts of early experimental work. The significance of hydrophobicity for the slip phenomenon does not seem to be
recognized at this early stage.

The matter was revisited at the beginning of 20th century as reviewed in ~\cite{vinogradova1999}. Several research groups reported faster flow after they had treated the capillary tube with organic compounds, and interpreted results in favour of slippage of water. There have been however some conclusions  about complications from surface tension effects in these early experiments. Although not recognized by the authors, their work
tried to make a connection between the amount of slip and the hydrophobization of the
solid.  That is why, despite the fact that the early results still did not provide a strong
support to the hypothesis of a hydrophobic slippage, we consider these papers to be an
important contribution to the subject. The first reliable results for water were probably obtained at the second part of 20th century. The measurements by Schnell~\cite{schnell.e:1956} and Churaev et al.~\cite{churaev.nv:1984} in thin hydrophobic capillaries unambiguously suggested a concept of hydrophobic slippage, which allowed to formulate first theoretical models~\cite{vinogradova.oi:1995a,vinogradova1999,barrat:99}. However, despite increasing body of evidence
in favor of hydrophobic slippage, experimental methods and results have been rather limited, the amplitude of slip and its dependence on various parameters (such as the contact angle, shear-rate, etc) still remained an open question.

In 21st century, or during the last decade, the field of hydrodynamic boundary conditions is rapidly advancing, motivated by potential applications in microfluidics, as well as fundamental scientific issues in colloid physics and fluid mechanics. Both hydrophobic and rough interfaces have received enormous attention. Note that some of the first data  completely escaped from
the theoretical framework, with qualitative (shear rate
threshold to slippage; shear rate dependent slip length)
and quantitative (micrometric versus nanometric slip
length) discrepancies,~\cite{lauga2005} which have caused hot debates. Fortunately, an avalanche development of the field, including experimental methods,  allowed one, and relatively fast, clarify the situation and highlight reasons for existing controversies. Thus, it is possible to conclude that boundary conditions at the smooth hydrophobic surfaces are currently quite well understood. We also suggest that a rough interface has recently been relatively well understood, although there are still some open questions. Over the past few years hot and promising directions of research seem to be the strategy of a combination of hydrophobic slip with surface roughness and/or interfacially driven flow, such as electro-osmosis. These new systems of interest could potentially lead to a generation of extremely fast flows in microfluidic devices and stimulated new fundamental studies.~\cite{vinogradova.oi:2010}

\section{Terminology and models}

\begin{figure}
\begin{center}
 \includegraphics [width=9 cm]{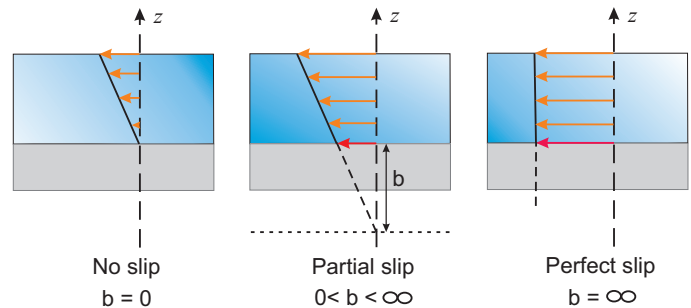}\,
   \end{center}
  \caption{Interpretation of the Navier slip length $b$.}
  \label{fig:slip_length0}
\end{figure}

 We will refer to as a slip any situation where the value of the tangential component of velocity appears to be different from that of the solid surface. The simplest possible relation assumes that the tangential force per unit area exerted on the solid surface is proportional to the slip velocity. Combining this with the constitutive equation for the bulk Newtonian fluid one gets the so-called (scalar) Navier boundary condition~\cite{navier1823}
\begin{equation}\label{slipBC}
u_s = b \frac{\partial u}{\partial z},
\end{equation}
where $u_s$ is the (tangential) slip velocity at the wall, $\partial u / \partial z$ the local shear rate,  and $b$ the slip length. This slip length represents a distance inside the solid to which the velocity has to be extrapolated to reach zero. The standard no-slip boundary condition corresponds to $b=0$, and the shear-free boundary condition corresponds to $b \to \infty$~\cite{vinogradova.oi:1995a} (see Fig.~\ref{fig:slip_length0}). In the most common situation $b$ is finite (a partial slip) and associated with the positive slip velocity. It can, however, be negative, although in this case it would not have a long-range effect on the flow~\cite{vinogradova.oi:1995d}.

Obviously the control of slip lengths is of major importance for flow at interface and in confined geometry. It would be useful to distinguish between three different situations for a boundary slip since the dynamics of fluids at the interface introduce various length scales.

\begin{figure}
\begin{center}
 (a)\includegraphics [width=7 cm]{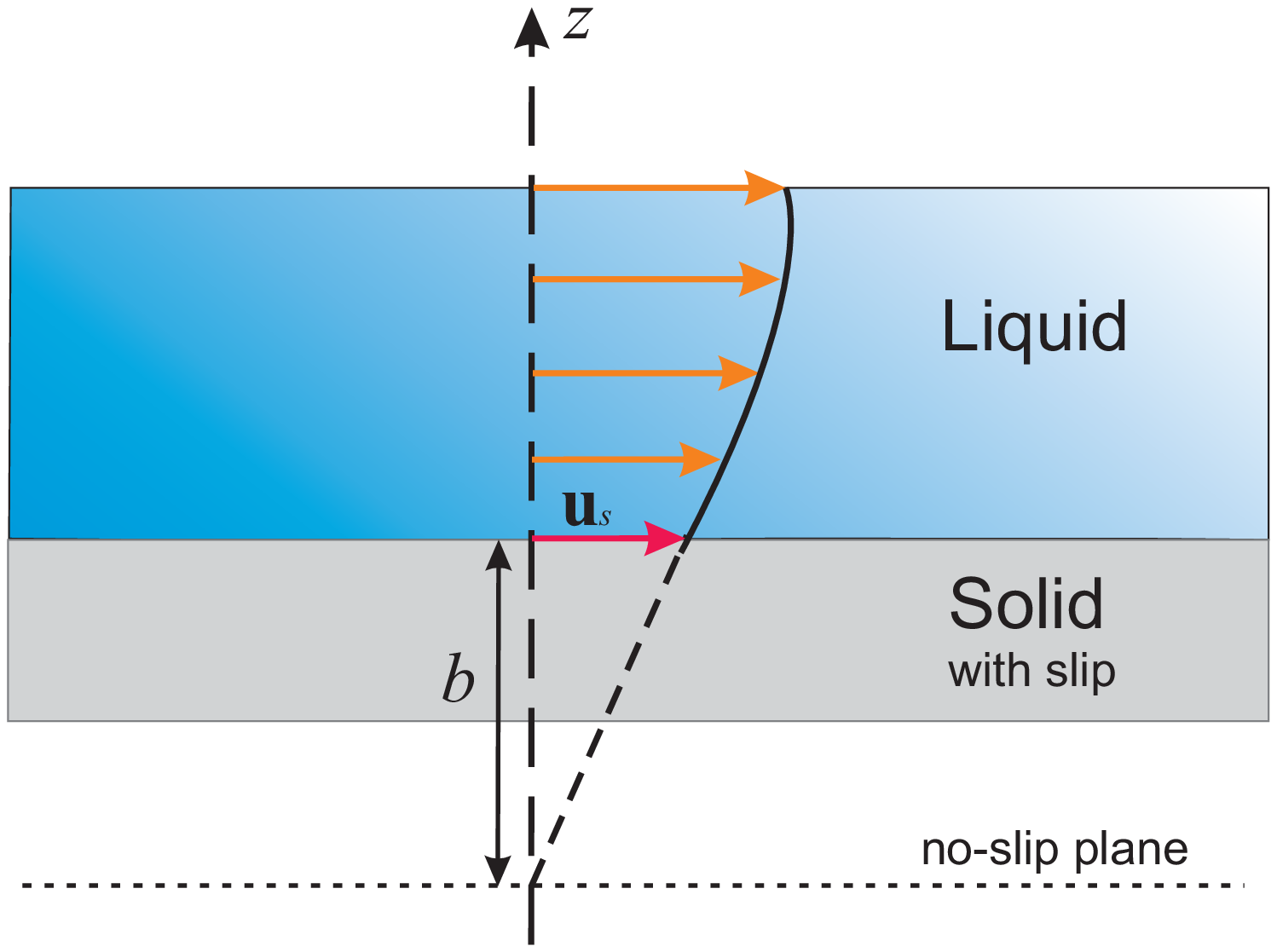}\,
 (b)\includegraphics [width=7 cm]{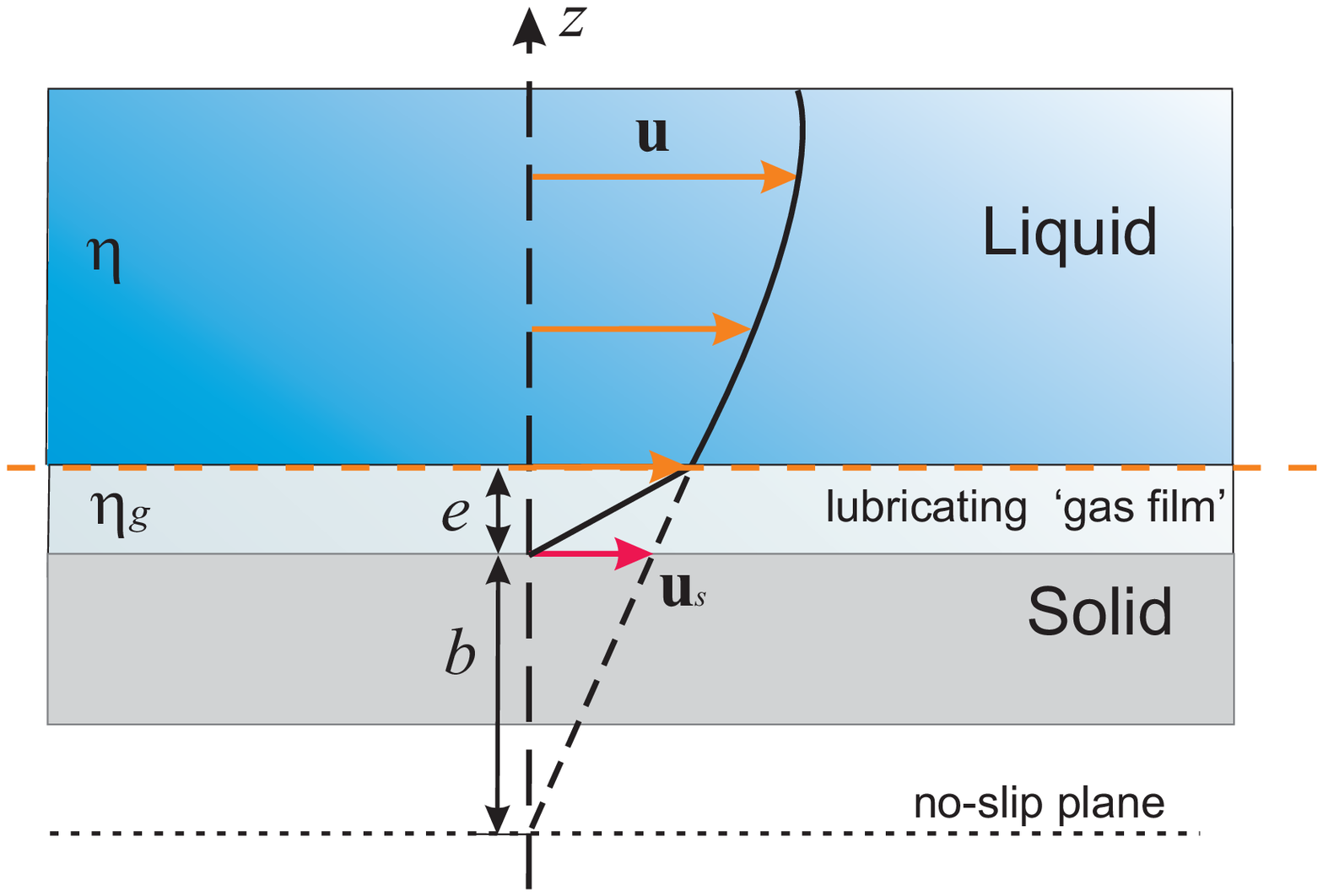}\\
 (c)\includegraphics [width=7 cm]{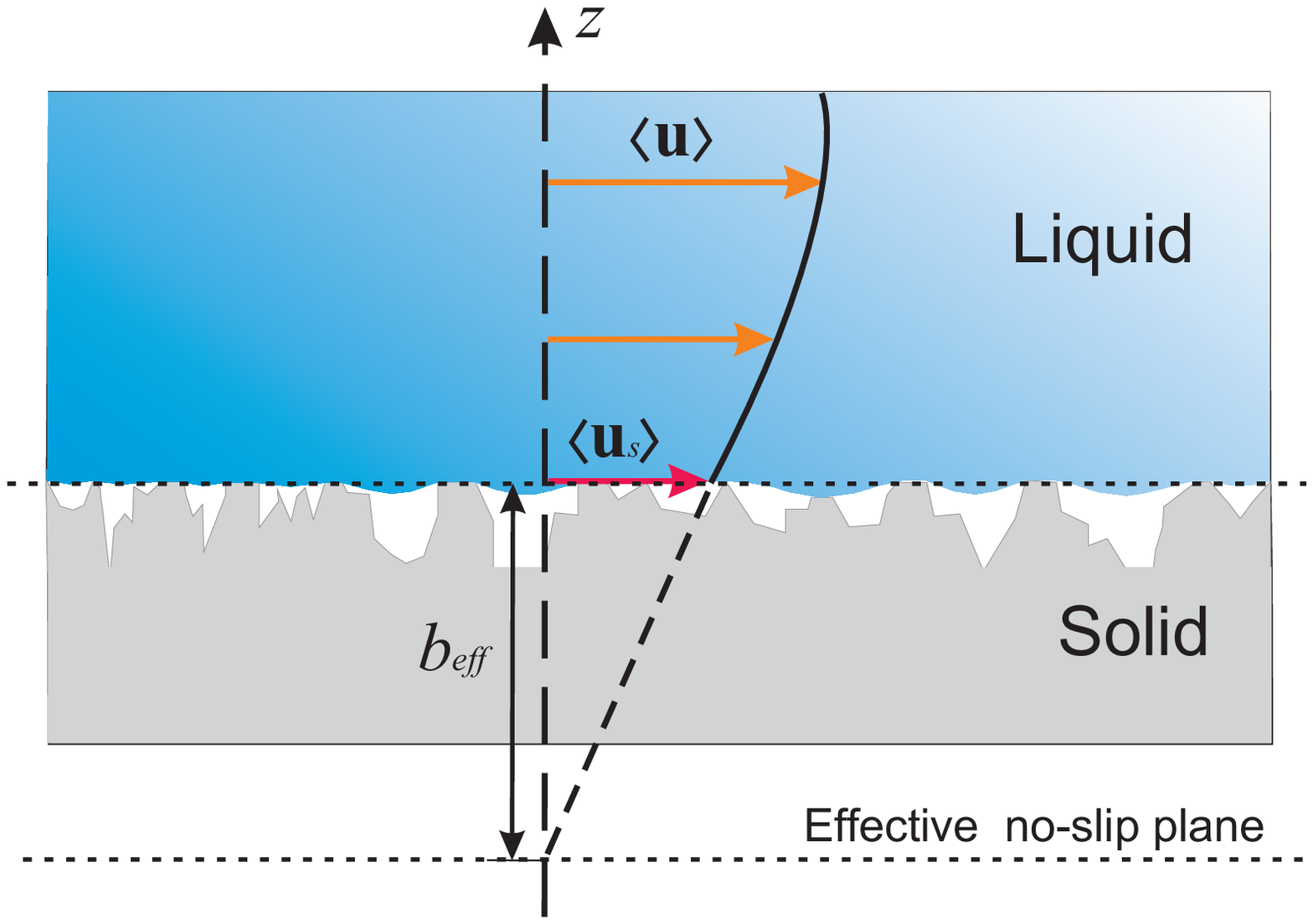}
  \end{center}
  \caption{Schematic representation of the definition of intrinsic (a), apparent (b), and effective (c) slip lengths. Reprinted with permission from~\cite{vinogradova.oi:2010}. Copyright (2011) by the IOP Publishing Ltd.}
  \label{fig:slip_length}
\end{figure}

 \p    \emph{Molecular} (or intrinsic) slip, which allows liquid molecules to slip directly over solid surface (Fig.~\ref{fig:slip_length}a). Such a situation is not of main concern here since molecular slip cannot lead to a large $b$~\cite{vinogradova1999,lauga2005,bocquet2007} and its calculations requires a molecular consideration of the interface region.

An important point to note that although Eq.(\ref{slipBC}) is the most commonly used boundary
condition for hydrodynamic slip, it is not widely appreciated that
\cite{navier1823} also postulated the more general relation, $ \Delta u = M \tau $
where $\tau$ is the local shear stress (normal traction) and $M$ is a
constant interfacial mobility (velocity per surface stress). For a
Newtonian fluid, $\tau = \eta \partial u /\partial z$, this reduces to
(\ref{slipBC}) with $b = M\eta$, where $\eta$ is the
viscosity. Molecular dynamics (MD) simulations have shown that
the equation with constant $M$ is more robust than
(\ref{slipBC}) with constant $b$, since the fluctuating slip
velocity correlates better with the shear stress (normal forces) than
with velocity gradients very close to the surface
\citep{bocquet2007}.

A possible starting point for molecular modeling is a scalar
Einstein relation, $M = b kT/A$, relating the $M$ to the
`interfacial diffusivity' per unit area $A$, by analogy with the
theory of Brownian motion. This can be recast as a type of a
Green-Kubo formula, and allows one to prove that the slip length does not depend on the bulk viscosity, $\eta$ (since it depends on $\eta D$).~\cite{bocquet2007} Rough  estimates of the molecular slip suggest $ b\propto k_B T \eta D /\epsilon^2$, where $\epsilon$ reflects a typical energy of interaction between liquid and solid. \cite{sendner.c:2009,bocquet2010} By relating  $\epsilon$ and a contact angle, $\cos\theta$ one can then propose a scaling relation $b\propto (1+\cos\theta)^{-2}$, which is consistent with MD results.\cite{huang.d:2008,sendner.c:2009} In other words, the static contact angle is the crucial parameter controlling water slippage over smooth hydrophobic surfaces.  We remark and stress, however, that for realistic contact angles only $b$ below 10 nm has been predicted.
Therefore, it is impossible to benefit of such a slip in a larger scale applications.

\p The intrinsic boundary condition maybe rather different from what is probed in flow experiment at larger length scale. It has been proposed~\cite{vinogradova.oi:1995a} to describe the interfacial region as a lubricating `gas film' of thickness $e$ of viscosity $\eta_g$ different from its bulk value $\eta$. A straightforward calculations give \emph{apparent} slip (Fig.~\ref{fig:slip_length}b)
\begin{equation}\label{apparent_slip}
b=e\left(\frac{\eta}{\eta_{\rm g}}- 1\right)\simeq e \frac{\eta}{\eta_{\rm g}}
\end{equation}
This represents the so-called `gas cushion model' of hydrophobic slippage~\cite{vinogradova.oi:1995a}, which got a clear microscopic foundation in terms of a prewetting transition~\cite{andrienko.d:2003}. Being a schematic representation of a depletion close to a wall~\cite{dammler.sm:2006}, this model provides a useful insight into the sensitivity of the interfacial transport to the structure of interface. Similarly, electrokinetic flow displays apparent slip. Note that recent molecular slip studies~\cite{huang.d:2008,sendner.c:2009} also suggested a kind of the `vapor cushion model', where  $b\propto e^4$, but apparently in a one-component system the value of $e$ is too small to describe most of experimental data, suggesting that a two-component system~\cite{vinogradova.oi:1995a,andrienko.d:2003} is required. Whether of not the addition of a hydrophobic solute will lead to a non-linear dependence of $b$ on $e$ remains an open question and has to be investigated.

A modification of this scenario would be a  nanobubble coated surface, i.e. a heterogeneous two-phase depletion layer.~\cite{vinogradova.oi:1995b,yakubov:00,borkent.b:2007,ishida:2000b,zhang2006, lohse_bubble2010}. However, as it has been discussed yet in~\cite{vinogradova1999} nanobubbles will reduce
the viscosity of the surface layer only  if surface tension is small enough (which means that the bubbles deform). If the shape of the nanobubbles will remain a spherical lens, $\eta_g$ should exceed the bulk viscosity. There
remains still an open question connected with the exact expression for the effective viscosity of the layer of
surface nanobubbles.

\p Another situation is that of \emph{effective} slip, $b_{\rm eff}$, which refers to a situation where slippage at a complex heterogeneous surface is evaluated by averaging of a flow over the length scale of the experimental configuration (e.g. a channel etc)~\cite{stone2004,Bazant08,feuillebois.f:2009,kamrin.k:2010}. In other words, rather than trying to solve equations of motion at the scale of the individual corrugation or pattern, it is appropriate to consider the  `macroscale' fluid motion (on the scale larger than the pattern characteristic length or the thickness of the channel) by using effective boundary conditions that can be applied at the imaginary smooth surface. Such an effective condition mimics the actual one along the true heterogeneous surface. It fully characterizes the flow at the real surface and can be used to solve complex hydrodynamic problems without tedious calculations. Such an approach is supported by a statistical diffusion arguments (being treated as an example of commonly used Onsager-Casimir relations for non-equilibrium linear response)~\cite{Bazant08}, theory of heterogeneous porous
materials~\cite{feuillebois.f:2009}, and has been justified for the case of Stokes flow over a broad class of surfaces~\cite{kamrin.k:2010}. For anisotropic textures $b_{\rm eff}$ depends on the flow direction and is generally a tensor~\cite{Bazant08}. Effective slip also depends on the interplay between typical length scales of the system as we will see below. Well-known examples of such a heterogeneous system include composite superhydrophobic (Cassie) surfaces, where a gas layer stabilized with a rough wall texture (Fig.~\ref{fig:slip_length}c). For these surfaces effective slip lengths are often very large compared with the value on flat solids, similarly to what has been observed for wetting, where the contact angle can be dramatically enhanced when surface is rough and heterogeneous~\cite{quere.d:2008}.

\section{Experimental methods}

The experimental challenge generated a considerable progress in experimental tools for investigating flow boundary conditions, using the most recent developments in optics and scanning probe techniques. A large variability still exists in
the results of slip experiments so it is important first
to consider the different experimental methods used to
measure slip.
Two broad classes of experimental approaches have been
used so far: indirect and direct (local) methods.

\begin{figure}
\begin{center}
\includegraphics[width=6cm,clip]{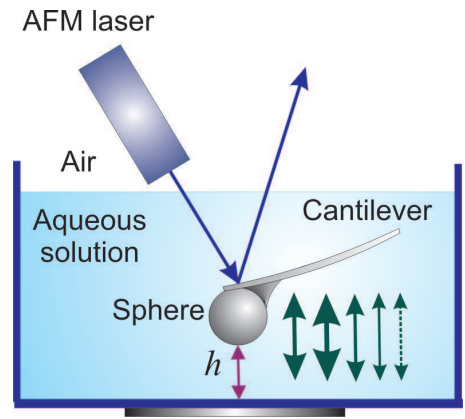}
\end{center}
\caption{Schematic of the dynamic AFM force experiment. Adapted from~\cite{vinogradova.oi:2006}} \label{fig:AFM}
\end{figure}

\p High-speed force measurements can be performed with
the SFA (surface forces apparatus)~\cite{chan.dyc:1985,charlaix.e:2005,horn:00}
or AFM (atomic force microscope)~\cite{vinogradova:03}. In particular, in the drainage method~\cite{chan.dyc:1985,vinogradova:03} the end of the spring away from the attached sphere is driven toward the
(fixed) plane with a constant driving speed (as shown in Fig.~\ref{fig:AFM}). The
sphere itself, however, does not move at a constant speed,
so that the spring is deflected as a result of both the surface force (which should be measured separately)
and the hydrodynamic forces. The solution of the (differential) equation of motion  allows to
deduce a drag force, with the subsequent comparison with a theory
of a film drainage,~\cite{vinogradova.oi:1995a,vinogradova:96} where the drag force is
\begin{equation}\label{drag_force}
    F=- \frac{6 \pi \eta R^2}{h} \frac{dh}{dt} f^{\ast} = F_T f^{\ast}
\end{equation}
where $- dh/dt$ is the relative velocity of the surfaces, $F_T$ is the Taylor solution for no-slip surfaces (which, however,
never appeared in any of G. I. Taylor's publications as
discussed in~\cite{horn:00}), and $f^{\ast}$ represents a correction function for slippage, which depends on $b/h$. For example, for a very convenient experimental configuration of the interaction of a hydrophilic surface with a hydrophobic one, which allows us to avoid the formation of a gas bridge,~\cite{andrienko.d:2004} the correction for slip takes the form~\cite{vinogradova.oi:1995d}
\begin{equation}\label{drag_fast}
    f^{\ast}= \frac{F}{F_T} = \frac{1}{4} \left(1+\frac{3 h}{2 b} \left[ \left(1 + \frac{h}{4 b}\right) \ln\left(1+\frac{4 b}{h}\right)-1 \right] \right)
\end{equation}
Note that when $h \ll 4b$, $ f^{\ast} \to 1/4$, so that the force is still inversely proportional to the gap, but becomes four times smaller than predicted for hydrophilic surfaces. For two similar hydrophobic surfaces, $f^{\ast} \to (h/3b) \ln (6b/h)$ when $h \ll 6b$, so that a hydrodynamic drag is only logarithmically dependent on the separation.~\cite{vinogradova.oi:1995a}

Note that  AFM force balance incorporates both (concentrated)
force on the sphere and the drag on the cantilever as shown in Fig.~\ref{fig:AFM}. The drag on a cantilever is neither
small nor negligible~\cite{vinogradova.oi:2006,vinogradova.oi:2001}, and its ignorance might cause wrong experimental conclusions.

This approach, being
extremely accurate at the nanoscale, does not provide
visualization of the flow profile, so that these
measurements are often identified as \emph{indirect}.

\begin{figure}
\begin{center}
\includegraphics[width=9cm,clip]{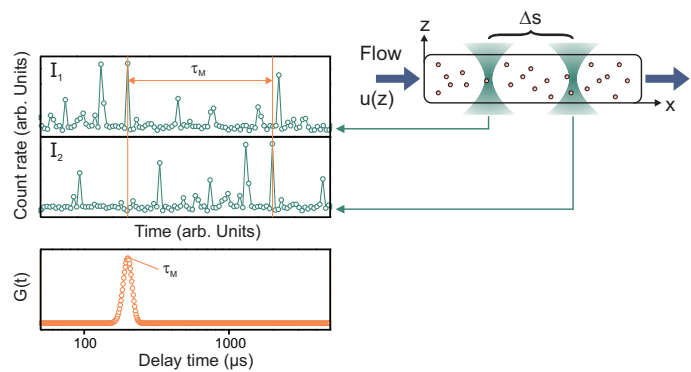}
\end{center}
\caption{Schematics of the double-focus spatial
fluorescence cross-correlation method. Two laser foci are placed
along the $x$ axis separated by a distance of a few $\mu$m. They
independently record the time-resolved fluorescence intensities
$I_1 (t)$ and $I_2 (t)$. The forward cross-correlation of these
two signals yields $G(t)$. Two foci are scanned simultaneously
along the $z$ axis to probe the velocity profile $u(z)$. Adapted from~\cite{lumma.d:2003,vinogradova.oi:2009}.} \label{fig:dffcs}
\end{figure}

\p \emph{Direct} approaches
to flow profiling, or velocimetry, take advantage of various
optics to monitor tracer
particles. These methods include TIR-FRAP (total internal reflection - fluorescence recovery after photo-bleaching)~\cite{pit:2000},
$\mu$-PIV~\cite{tretheway.dc:2002,josef.p:2005} (particle image velocimetry),
TIRV (total internal reflection velocimetry)~\cite{huang.p:2006}, EW $\mu$-PIV(evanescent wave micro particle image velocimetry)~\cite{zettner2003}, and multilayer nano-particle image velocimetry (nPIV)~\cite{li2006, maynes_turb1}. Their
accuracy is normally much lower than that of force methods due to
relatively low optical resolution, system noise due to
polydispersity of tracers, and difficulties in decoupling
flow from diffusion (the tracer distribution in the flow field is affected by Taylor dispersion~\cite{vinogradova.oi:2009}). As a consequence, it has been always
expected that a slippage of the order of a few tens nanometers
cannot be detected by a velocimetry technique. However, recently direct high-precision measurements at the
nanoscale have been performed with a new optical technique, based on a DF-FCS (double-focus spatial fluorescence
cross-correlation)~\cite{lumma.d:2003,vinogradova.oi:2009} (as is schematically shown in
Fig.~\ref{fig:dffcs}).  As the fluorescence tracers are flowing along the
channel they are crossing consecutively the two foci, producing
two time-resolved fluorescence intensities $I_1(t)$ and $I_2(t)$
recorded independently. The time cross-correlation function can be calculated and
typically exhibits a local maximum. The position of this maximum
$\tau_{\rm M}$ is characteristic of the local velocity of the
tracers. Another example of high resolution promising applications of FCS consists in determination of average transverse diffusion coefficient to probe slippage~\cite{joly.l:2006}.  Since FCS methods allow consideration of $N \sim 10^6$ particles, this gives a satisfactory signal to noise ratio $\sqrt{N}$ of order $10^3$, providing extremely good resolution as compared with other direct velocimetry methods. Coupling with TIRF~\cite{yordanov.s:2009}, which  allows the measurements of the distance of tracers from the wall through the exponential decay of an evanescent wave, should further improve the accuracy of approach.

\begin{figure}
\begin{center}
\includegraphics[width=7cm,clip]{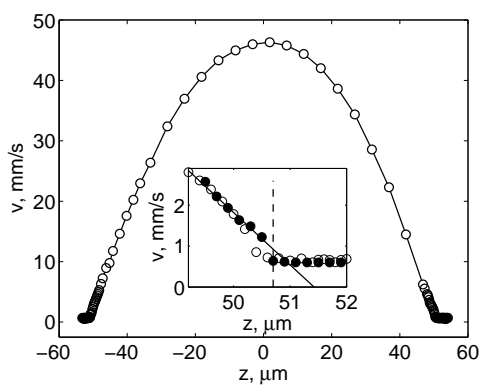}
\end{center}
\caption{Typical velocity profile $v(z)$ (open circles) measured in a $\sim 100$
$\mu$m channel with $10^{-4}$ mol/L NaCl solution. Inset shows the same profile in the vicinity of the wall. Error bars were determined from the deviation among repeated measurements. When error bars are absent, uncertainties
are smaller than the circles. Dots are simulation results. Dash-dotted line fits data (the apparent
slip length, $b_{\rm app} \sim 740$ nm), and dashed line shows the wall location
($z=50.69 \mu$m). Adapted from~\cite{vinogradova.oi:2009}} \label{fig:profile}
\end{figure}

In the vicinity of the wall the tracers are
submitted to a Taylor dispersion, i.e. their diffusion combined
with shear enhances the migration speed in the flow direction. This phenomenon seriously complicates all velocimetry methods since to extract the velocity of fluid this effect should be modelled precisely. As shown by analysis of DF-FCS data,~\cite{vinogradova.oi:2009} large observed values of the apparent slip at the hydrophilic wall are normally fully attributed to a Taylor
dispersion of nanotracers (see Fig.~\ref{fig:profile}). The data obtained with other velocimetry technique still await clarification. We suggest however that some very large values of a hydrophobic slip~\cite{tretheway.dc:2002,pit:2000} might reflect a Taylor dispersion too.

Many experiments have been performed on the subject, with somewhat contradicting  results. Experimental work focussed mostly on bare (smooth) surface, more recent investigations have turned towards rough and structured surfaces, in particular super-hydrophobic surfaces~\cite{rothstein.jp:2010}. We refer the reader to comprehensive review articles~\cite{lauga2005,neto.c:2005} for detailed account of early experimental work. In our chapter we mention only what we believe is the most relevant recent contribution to the subject of flow past `simple' smooth hydrophobic and rough hydrophilic surfaces, which clarified the existing controversies in the field. We focus, however, more on the implication of micro- and nanostructuring on fluidic transport, which is still at its infancy and remains to be explored.

\section{Smooth surfaces: Slippage vs wetting}

From the theoretical~\cite{bocquet2007,bocquet2010} and simulation~\cite{sendner.c:2009,kunert.c:2010}  point of view slippage should not appear on a hydrophilic surface, except probably
as at very high shear rate~\cite{thompson.pa:1997}. A slip length
of the order of hundred nanometers or smaller is, however,
expected for a hydrophobic
surface~\cite{vinogradova.oi:1995a,bocquet2007,andrienko.d:2003,bib:jens-kunert-herrmann:2005}.

On the
experimental side, no consensus was achieved until recently. While some
experimental data were consistent with the theoretical expectations
both for
hydrophilic and hydrophobic
surfaces ~\cite{charlaix.e:2005,vinogradova:03,li.h:2010},
some other reports completely escaped from this picture with both
quantitative (slippage over hydrophilic surface, shear rate
dependent slippage, rate threshold for slip, etc) and quantitative
(slip length of several $\mu$ms) discrepancies (for a
review see~\cite{lauga2005}).
More recent experiments, performed with various new experimental methods, finally concluded that water does not slip on smooth hydrophilic surfaces, and develops a slip only on hydrophobic surface~\cite{vinogradova.oi:2009,joly.l:2006,vinogradova.oi:2006,honig.cdf:2007,bouzigues.c:2008,maali.a:2006}. One can therefore conclude that a concept of hydrophobic slippage is now widely
accepted.

An important issue is the amplitude of hydrophobic slip. The observed slip length reached the range 20-100 nm, which is above predictions of the models of molecular slip~\cite{huang.d:2008,barrat:99}. This suggests the apparent slip, such as the `gas cushion model', Eq.~(\ref{apparent_slip}). Water glides on air, owing to the large
viscosity ratio between water and air (typically a factor of 50). Experimental values of $b$ suggest that the thickness of this `layer' is below 2 nm. Another scenario of apparent slip such as a  nanobubble coated surface~\cite{vinogradova.oi:1995b,yakubov:00,borkent.b:2007,ishida:2000b} has to be explored in more details.

An important conclusion is that it is impossible to benefit of such a nanometric slip at separations O($\mu$m) and larger, i.e. in microfluidic applications. This is why in the discussion of super-hydrophobic slippage below we often ignore a slip past hydrophobic solids. However, a hydrophobic slippage is likely of major importance in nanochannels (highly confined hydrophobic pores, biochannels, etc), where ordinary Poiseuille flow is fully suppressed.

\section{Rough surfaces}

\begin{figure}
  \begin{center}
  (a)\includegraphics[width=0.41\textwidth, trim=0 -20 0 0]{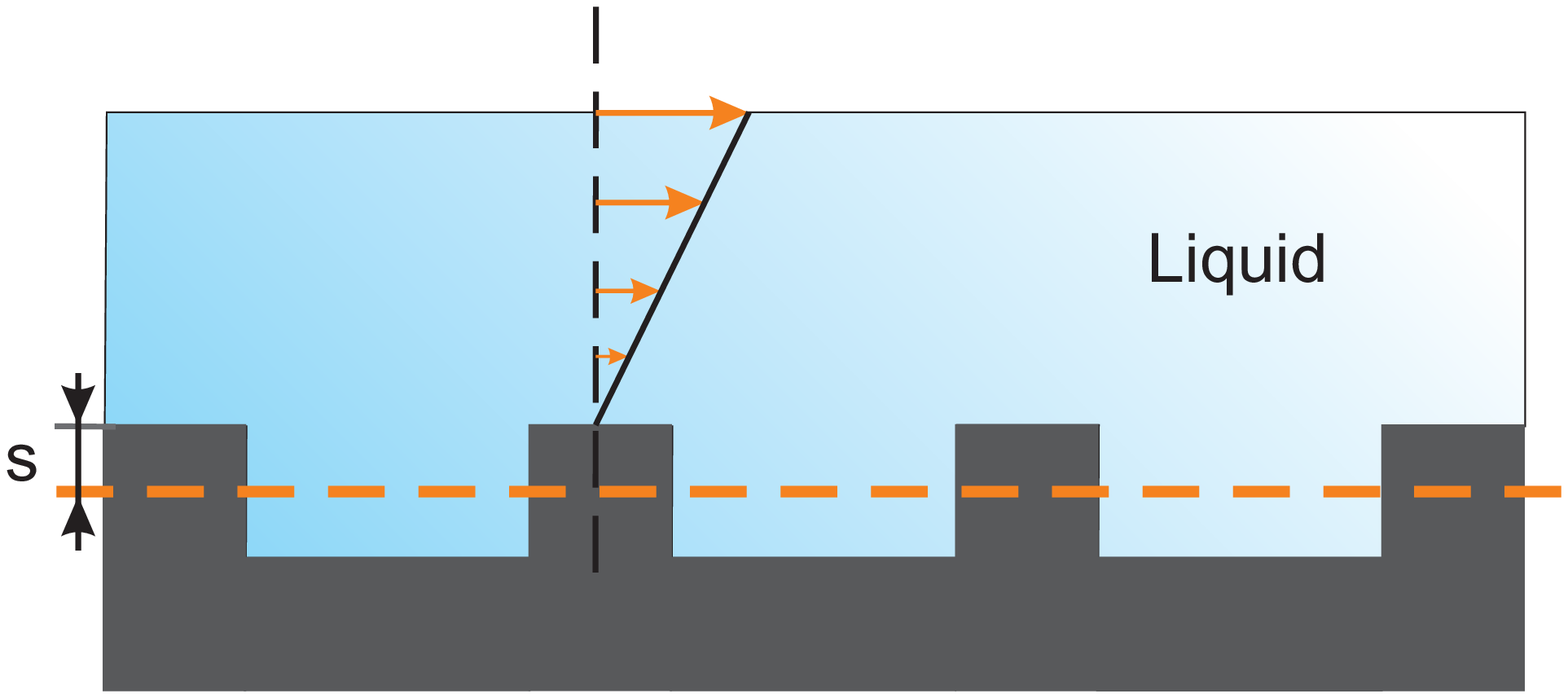}\,\,
  (b)\includegraphics[width=0.41\textwidth, trim=-20 20 0 0]{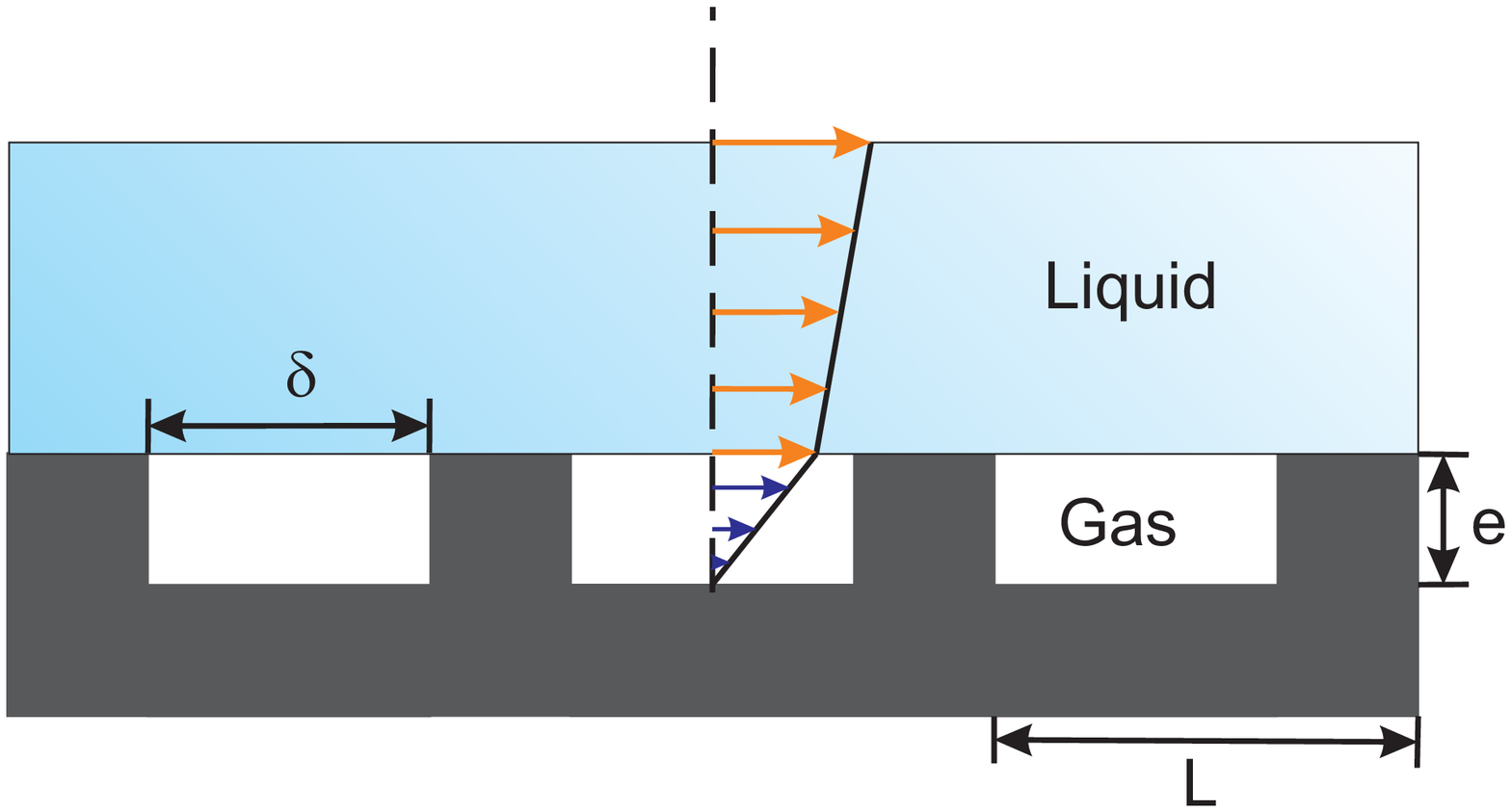}
  \end{center}
    \caption{Schematic representation of the (a) Wenzel and (b) Cassie pictures with the local flow profiles at the gas and solid areas. Reprinted with permission from~\cite{vinogradova.oi:2010}. Copyright (2011) by the IOP Publishing Ltd. }\label{fig:model}
\end{figure}

\begin{figure}
  \begin{center}
  (a)\includegraphics[width=0.6\textwidth]{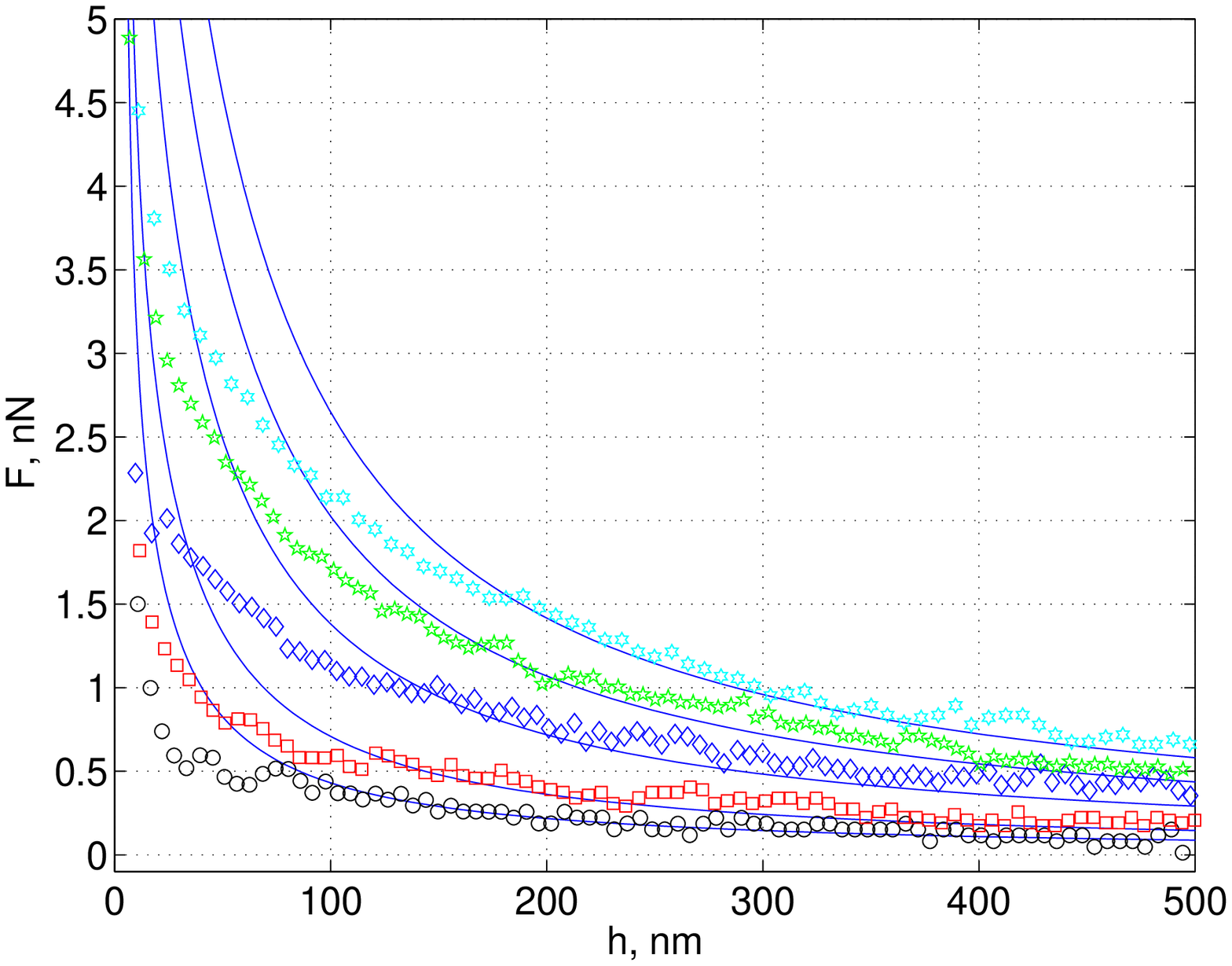}\\
  (b) \includegraphics[width=0.5\textwidth]{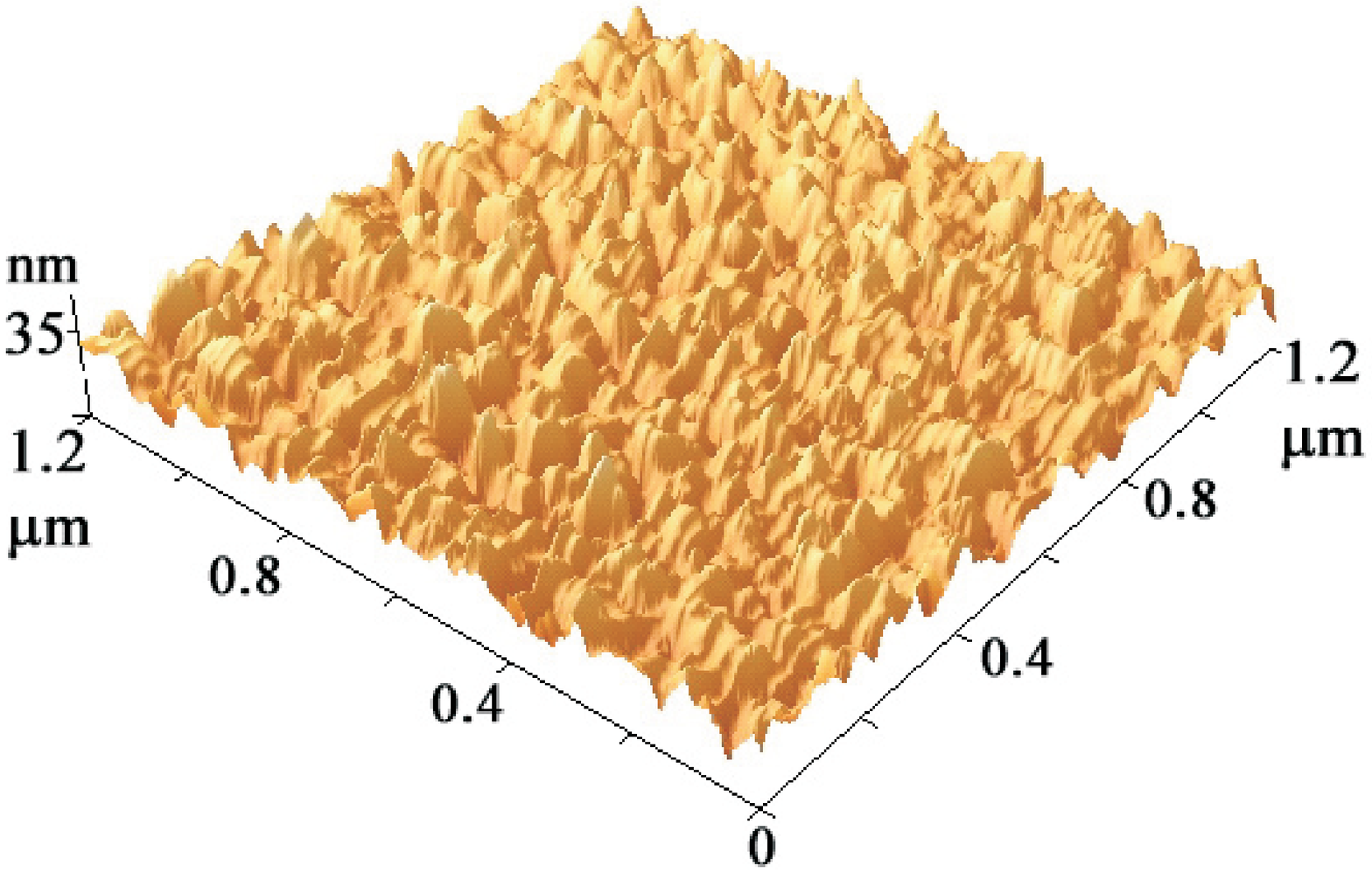}
    \end{center}
    \caption{(a) Hydrodynamic force acting on a rough sphere ($R=20 \mu$m) approaching a smooth plane. From
bottom to top are data (symbols) obtained at the driving speed -6,
-10, -20, -30, and -40 $\mu$m/s. Solid curves show the calculation
results obtained for the same speed, by assuming no-slip boundary conditions. (b) An AFM image of an apex of a gold-coated
sphere. Adapted from~\cite{vinogradova.oi:2006}}\label{fig:afm_roughness}
\end{figure}

Only a very few solids are molecularly smooth. Most of them are naturally rough, often at a micro- and nanoscale, due to their structure, methods of preparation, various coatings. These surfaces are very often in the Wenzel (impaled) state, where solid/liquid interface has the same area as the solid surface (Fig.~\ref{fig:model}a). However, even for rough hydrophilic Wenzel surfaces the situation was not very clear, and
opposite experimental conclusions have been made: one is that roughness
generates extremely large slip~\cite{bonaccurso.e:2003}, and one is
that it decreases the degree of slippage~\cite{granick.s:2003,granick:02, maynes_3}.
A conclusion in favor of slip past rough surfaces is often made based on the fact that the hydrodynamic drag force becomes smaller when one of the surfaces is rough (see Fig.\ref{fig:afm_roughness}), which is qualitatively consistent with the model of slip, Eqs.(\ref{drag_force}),(\ref{drag_fast}).
Recent experimental data suggested that the description of flow near rough
surfaces has to be corrected. Such a flow is equivalent to expected near a smooth hydrophilic (i.e. no-slip) surface that is located between top and bottom of asperities~\cite{vinogradova.oi:2006}. In other words the correction should be done not for a slip, but for a separation from a rough surface. It has also been shown that opposing conclusions made by~\cite{bonaccurso.e:2003,granick.s:2003,granick:02} simply reflect a different
way of a definition of zero separation in the AFM (at the top of asperities) and SFA (at their
bottom).

The theoretical description of such a flow
represents a very difficult problem. Previous
theoretical investigations addressed only a case of  a far-field flow,
where spatial variations are small compared to the typical length scale of the problem.
Most of the articles considered the issue of periodic roughness~\cite{lecoq.n:2004,jaeger.w:2003},
although some recent work exploited non-periodic, spatially homogeneous random roughness~\cite{basson.a:2008}.
It has been concluded that these situations can be modeled by the application of the Navier condition to an equivalent
smooth plane deposited on the top of roughness~\cite{lecoq.n:2004,kunert-harting-07}.  However, there have been also theoretical (far-field) arguments that
this should be rather a smooth equivalent no-slip wall placed at some distance below the top, but still above the bottom
of
roughness~\cite{bechert.dw:1989}.

\begin{figure}
\begin{center}
\includegraphics[width=0.6\textwidth]{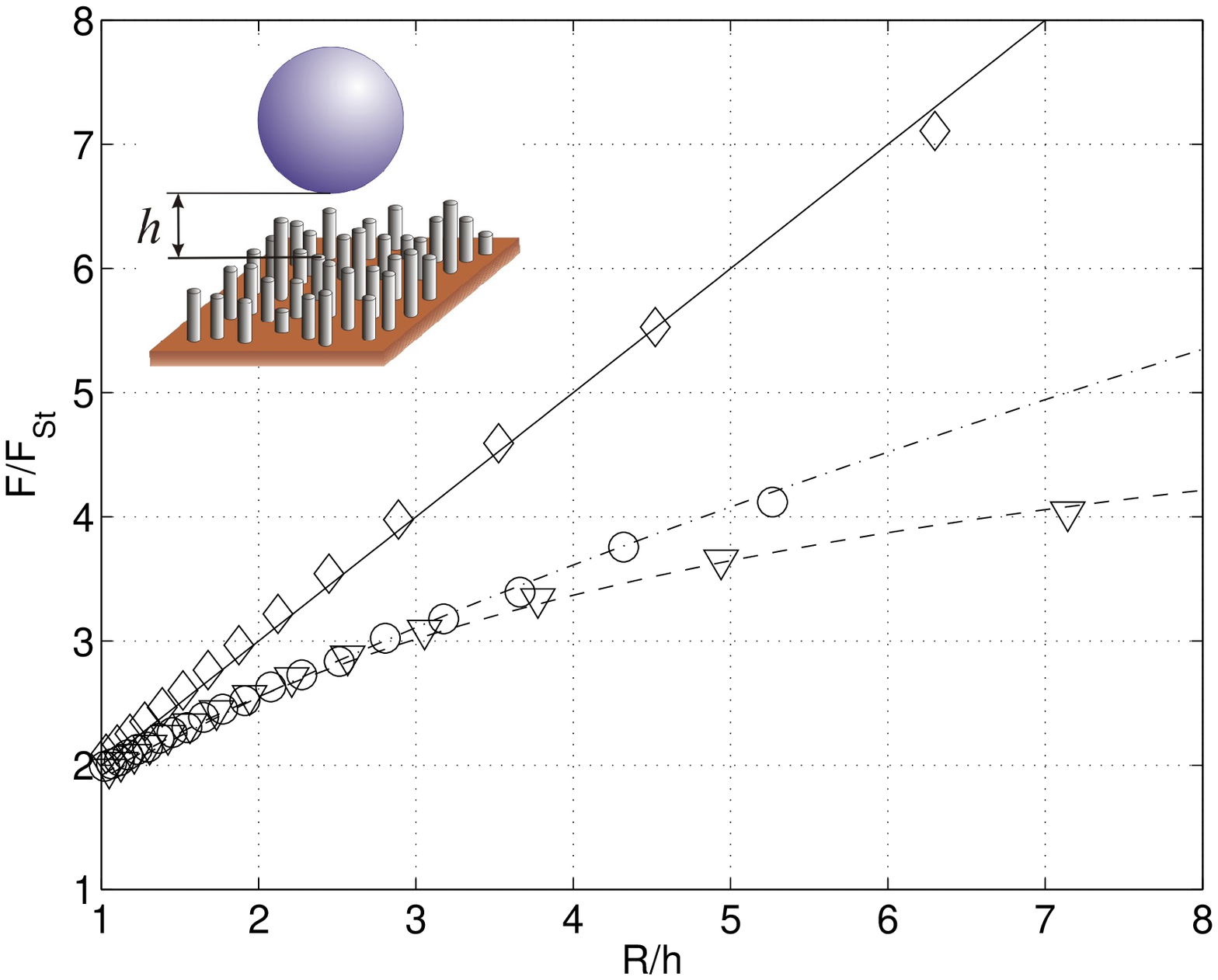}
\end{center}
\caption{Hydrodynamic force acting on a hydrophilic sphere of radius $R$ approaching a smooth hydrophilic (diamonds), smooth hydrophobic (circles) and randomly rough hydrophilic (triangles) wall with $\phi_2=4\%$ (adapted from~\cite{kunert.c:2010}). Here $F_{St} = 6 \pi \eta R U$ is the Stokes drag. The separation $h$ is defined on top of the surface roughness as shown in the inset.   Simulation results (symbols) compared with theoretical curves: $F/F_{St}=1+9 R/(8 h)$ (solid), $F/F_{St}=1+ 9 R f^{\ast}/(8 h)$ with $f^{\ast}=f(b/h)$, calculated with Eq.(\ref{drag_fast}) (dash-dotted), and $F/F_{St}=1+ 9 R/(8 [h+s])$ (dashed). Values of $b$ and $s$ were determined by fitting the simulation data.}
\label{fig:picture}
\end{figure}

This issue was recently resolved in the LB (lattice Boltzmann) simulation study~\cite{kunert.c:2010} where
the hydrodynamic interaction between a smooth sphere of radius $R$ and a
randomly rough plane was studied (as shown in Fig.~\ref{fig:picture}). Beside its
significance as a geometry of SFA/AFM dynamic force experiments, this
allowed one to explore both far and near-field flows in a single
`experiment'. The `measured' hydrodynamic force was smaller than predicted for two smooth
surfaces (with the separation defined at the top of asperities) if the standard no-slip boundary conditions are used in the
calculation. Moreover, at small separations
the force was even weaker and shows different asymptotics than expected if
one invokes slippage at the smooth fluid-solid interfaces. This can only be explained by the model of a no-slip wall, located at an intermediate position
(controlled by the density of roughness elements) between top and bottom
of asperities (illustrated by dashed line Fig.~\ref{fig:model}a). Calculations based on this model provided an excellent
description of the simulation data (Fig.~\ref{fig:picture}).

\section{Super-hydrophobicity and effective hydrodynamic slippage}\label{sec:super}

On hydrophobic solids, the situation is different from that on hydrophilic solids. If the solid is rough enough, we do not expect that the liquid will conform to the solid surface, as assumed in the Wenzel or impaled state. Rather air pockets should form below the liquid, provided that the energetic cost associated with all the
corresponding liquid/vapor interfaces is smaller than the energy gained not to follow the solid~\cite{quere.d:2005}. This is so-called Cassie or fakir state. Hydrophobic Cassie materials generate large contact angles and small hysteresis, ideal conditions for making water drops very mobile. It is natural to expect a large effective slip in a Cassie situation. Indeed, taking into account that the variation of the texture height, $e$, is in the typical interval $0.1-10$ $\mu$m, according to Eq.(\ref{apparent_slip}) we get $b=5-500$ $\mu$m at the gas area. The composite nature of the texture requires
regions of very low slip (or no slip) in direct contact with the liquid,
so the effective slip length of the surface, $b_{\rm eff}$, is smaller than $b$. Still, one can expect that a rational design of such a texture could lead to a large values of $b_{\rm eff}$. Below we make these arguments more
quantitative.

We examine an idealized
super-hydrophobic surface in the Cassie state sketched in Fig.~\ref{fig:model}b
where a liquid slab lies on top of the surface
roughness. The liquid/gas interface is assumed to be flat with no meniscus
curvature, so that the modeled super-hydrophobic surface
appears as a perfectly smooth with a pattern of
boundary conditions. In the simplified description the latter are taken as no-slip ($b_1=0$) over solid/liquid
areas and as partial slip ($b_2=b$) over gas/liquid regions [as we have shown above, $b_1$ is of the orders of tens nm, so that one could neglect it since $b_2$ is of the order of tens of $\mu$m]. We denote
as $\delta$ a the typical length
scale of gas/liquid areas. The fraction of solid/liquid
areas will be denoted $\phi_1=(L-\delta)/L$, and of gas/liquid area $\phi_2=1-\phi_1=\delta/L$.
Overall, the description of a super-hydrophobic surface we use here is similar to
those considered in Refs~\cite{feuillebois.f:2009,ybert.c:2007,belyaev.av:2010a,cottin.c:2004,cottin_bizonne.c:2003.a,priezjev.nv:2005, maynes1, maynes2}.
In this idealization, some assumptions may have a possible influence
on the friction properties and, therefore, a hydrodynamic force. First, by assuming flat
interface, we have neglected an additional mechanism for a dissipation connected with the meniscus curvature~\cite{harting.j:2008,lauga2009,sbragaglia.m:2007}. Second, we ignore a possible transition towards impaled (Wenzel) state that can be provoked by additional pressure in the liquid phase~\cite{pirat.c:2008,reyssat.m:2008}. Third, we do not take into account the circulation of air in the gas phase~\cite{ng.co:2010, maynes1, maynes2} and some possible other effects such as a depinning of the contact line~\cite{gao2009}.

Finally, for the sake of brevity we focus below only on the canonical microfluidic geometry where the fluid is confined between flat plates, and only on the asymmetric case, where one (upper) surface is smooth hydrophilic  and another (lower) represents a super-hydrophobic wall in the Cassie state. Such a configuration is relevant for various setups, where the alignment of opposite textures is inconvenient or difficult. We also restrict the discussion by a pressure-driven flow governed by the Stokes equations:
\begin{equation}\label{Stokes}
   \eta\nabla^2\textbf{u}=\nabla p,\,\,\,
  \nabla\cdot\textbf{u}=0,
\end{equation}
where $\textbf{u}$ is the velocity vector, and $p$ is pressure. Extensions of our analysis to study other configuration geometries and types of flow
would be straightforward.

\subsection{Anisotropic surfaces.}

Many natural and synthetic textures are isotropic. However,
it can be interesting to design directional structures, such as arrays of parallel grooves or microwrinkles,
that consequently generate anisotropic effective slip in the Cassie regime. The hydrodynamic slippage
is quite different along and perpendicular to the grooves. Axial motion is preferred, and
such designs are appropriate when liquid must be guided.
There are examples of such patterns in nature, such as the wings of butterflies or water striders.

\begin{figure}
\begin{center}
  \includegraphics [width=7.5 cm]{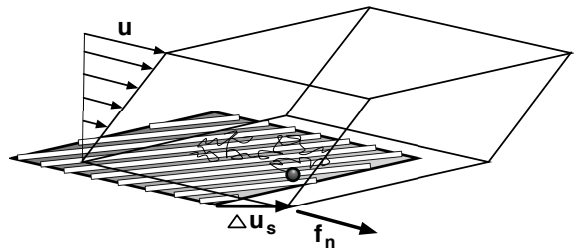}
  \end{center}
\caption{Sketch of tensorial hydrodynamic slip. The
  normal traction $\fb_n$ exerted by the fluid on an anisotropic
  surface produces an effective slip velocity $\Delta\ub=\Mb \, \fb_n$
  in a different direction. At the molecular level, the interfacial
  mobility tensor $\Mb$ is related to the trajectories of diffusing
  particles in the interfacial region, such as the one shown. Reprinted with permission from~\cite{Bazant08}. Copyright (2008) by the Cambridge University Press.  }
\label{fig:slab}
\end{figure}

\begin{figure}
\begin{center}
  \includegraphics [width=7.5 cm]{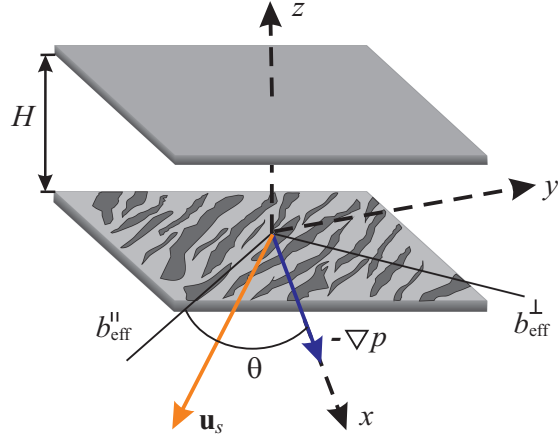}
  \end{center}
\caption{Sketch of a flat channel of thickness $H$ wall with notation for directions along the plates. One wall represents an anisotropic super-hydrophobic texture. Adapted from~\cite{vinogradova.oi:2010}}
\label{fig:tensor}
\end{figure}

A generalization of the Navier mobility condition is~\cite{Bazant08}
\begin{equation}
\Delta \ub  = \Mb \cdot \, (\nhat \cdot \sigmab) \label{eq:M}
\end{equation}
where $\fb_n = \nhat \cdot \sigmab$ is the fluid force (normal
traction) on the interface, $\sigmab$ is the local stress tensor, and
$\Mb$ is an interfacial mobility tensor. As shown in
Figure~\ref{fig:slab}, the effective slip vector is generally
misaligned with the force vector for an anisotropic surface, which has been analyzed in a number of studies~\cite{ajdari2002,stroock2002b}.  For anisotropic surfaces, the
mobility must be a second-rank tensor $\Mb = \{M_{ij}\}$. In the case of
a Newtonian fluid,  $\bb_{\rm eff} = \Mb \eta$, and we get a tensorial version of (\ref{slipBC}), as discussed in~\cite{stone2004,Bazant08}
\begin{equation}\label{effBC}
  \langle u_i |_A \rangle = \sum_{j,k} b^{\rm eff}_{ij} n_k \left\langle \left. \frac{\partial u_j}{\partial x_k} \right|_A \right\rangle,
\end{equation}
where $\langle\textbf{u}|_A\rangle$ is the effective slip velocity, averaged over the surface pattern and  $\textbf{n}$ is a unit vector normal to the surface $A$. The second-rank effective slip tensor $\textbf{b}_{\rm eff}\equiv\{b^{\rm eff}_{ij}\}$ characterizes the surface anisotropy and is represented by symmetric, positive definite $2\times 2$ matrix diagonalized by a rotation:
\begin{equation}
{\bf b}_{\rm eff}= {\bf S}_{\theta} \left(
\begin{array}{cc}
 b^{\parallel}_{\rm eff} & 0\\
 0 & b^{\perp}_{\rm eff}
\end{array}
\right) {\bf S}_{-\theta} ,
\qquad
{\bf S}_{\theta}=
\left(
\begin{array}{cc}
 \cos~\theta &  \sin~\theta \\
 -\sin~\theta & \cos~\theta
\end{array}
\right). \label{Tensor_Rotation}
\end{equation}
As proven in~\cite{Bazant08} for all anisotropic surfaces the eigenvalues $b^{\parallel}_{\rm eff}$ and $b^{\perp}_{\rm eff}$ of the slip-length tensor correspond to the fastest (greatest forward slip) and slowest (least forward slip) directions, which are always orthogonal (see Fig.~\ref{fig:tensor}).

\begin{figure}
\begin{center}
(a)\includegraphics*[width=0.17\textwidth]{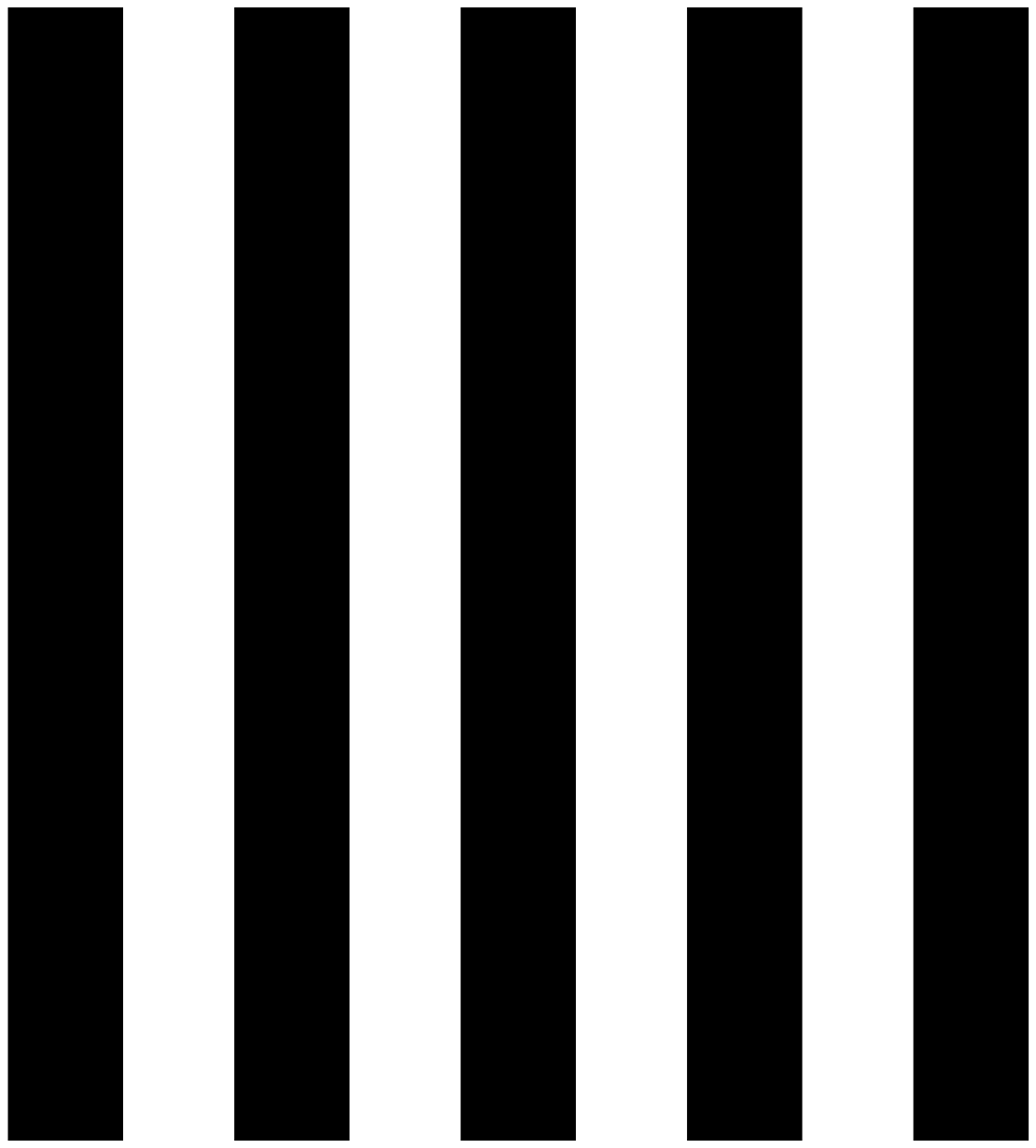}
(b)\includegraphics*[width=0.17\textwidth]{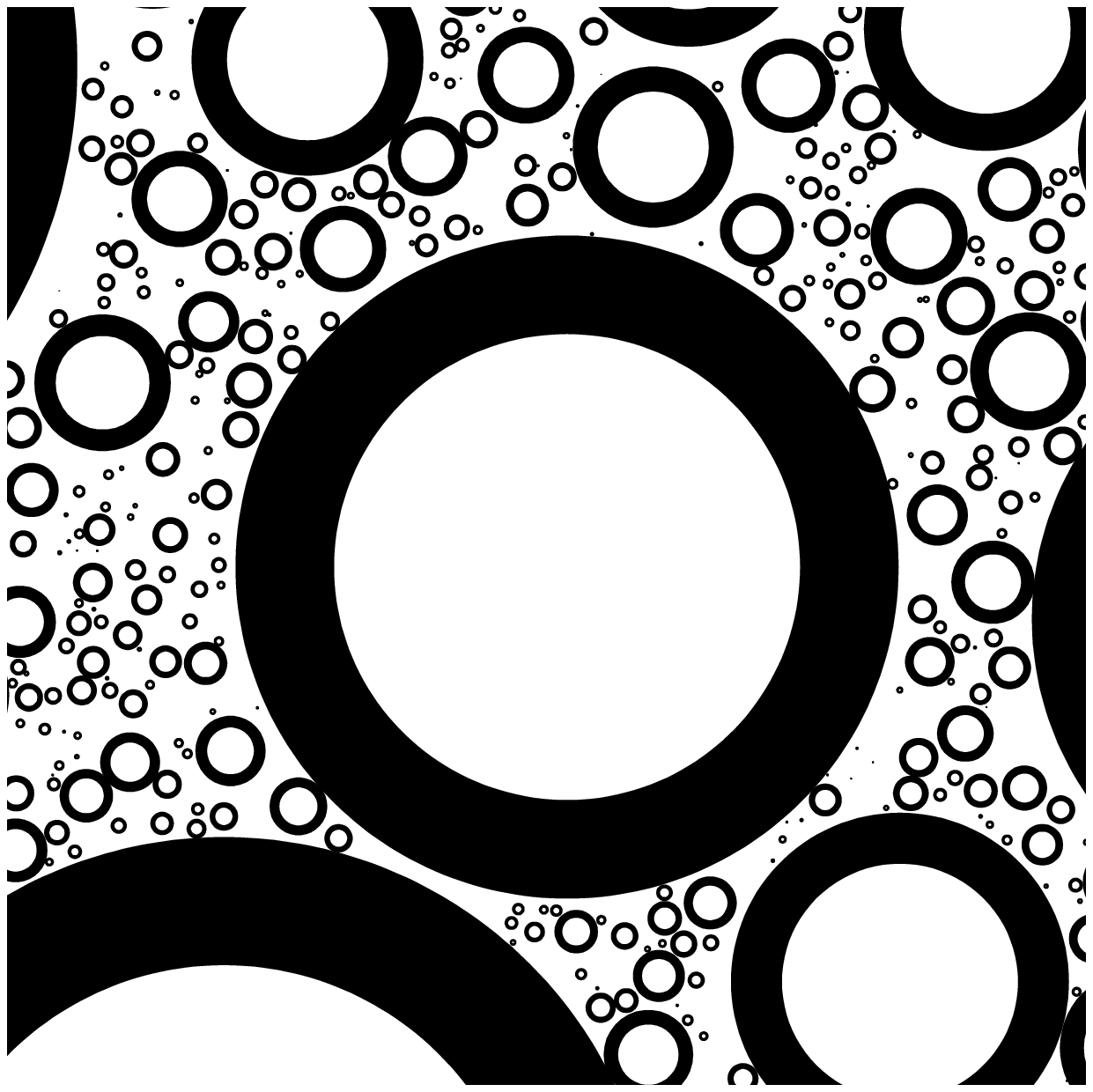}
(c)\includegraphics*[width=0.17\textwidth]{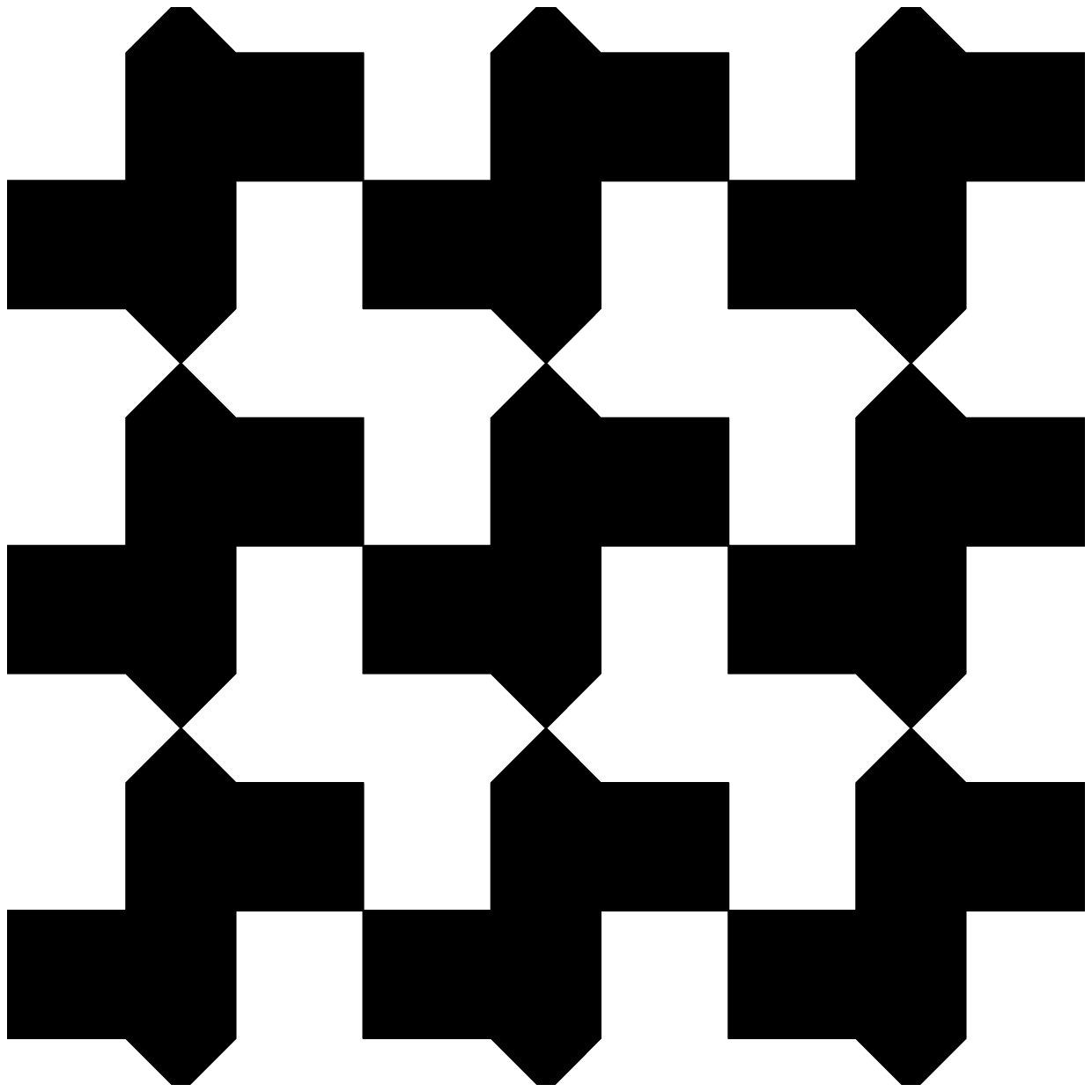}
(d)\includegraphics*[width=0.17\textwidth]{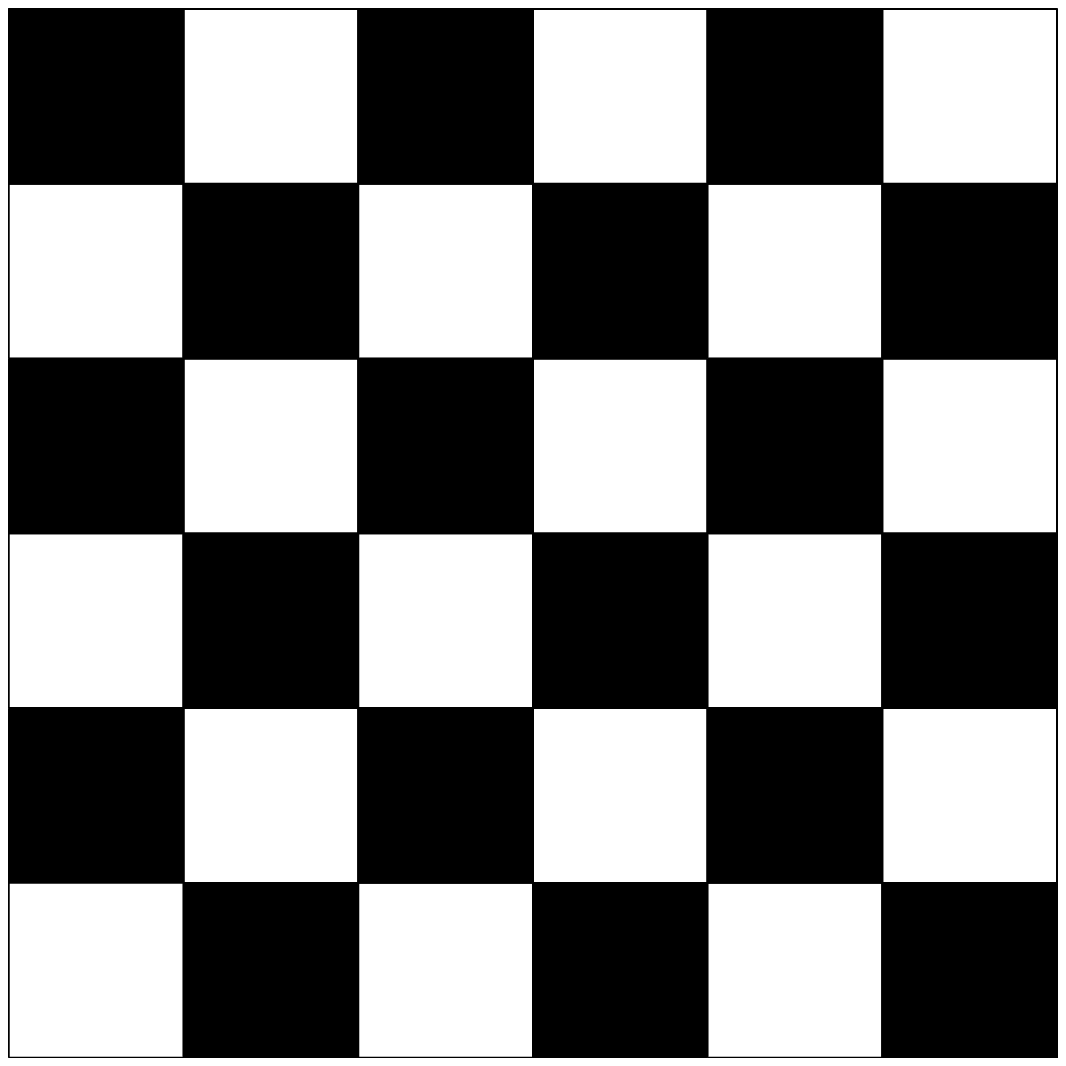}
\end{center}
  \caption{Special textures arising in the theory: (a)
  stripes, which attain the Wiener bounds of maximal and
  minimal effective slip, if oriented parallel or perpendicular to the
  pressure gradient, respectively; (b) the Hashin-Shtrikman
  fractal pattern of nested circles, which attains the maximal/minimal slip among all
  isotropic textures (patched should fill up the whole space, but their number is limited here for clarity); and (c) the Schulgasser and (d) chessboard textures, whose
  effective slip follows from the phase-interchange
  theorem. Adapted from~\cite{belyaev.av:2010b}}
  \label{fig:designs}
\end{figure}

To illustrate the calculation of the slip-length tensor, below we consider the geometry where the liquid is confined between two plates separated by a distance $H$, and one of them represents a super-hydrophobic striped wall (Fig.~\ref{fig:designs}a).

\begin{figure}
\begin{center}
(a) \includegraphics [width=0.45\textwidth]{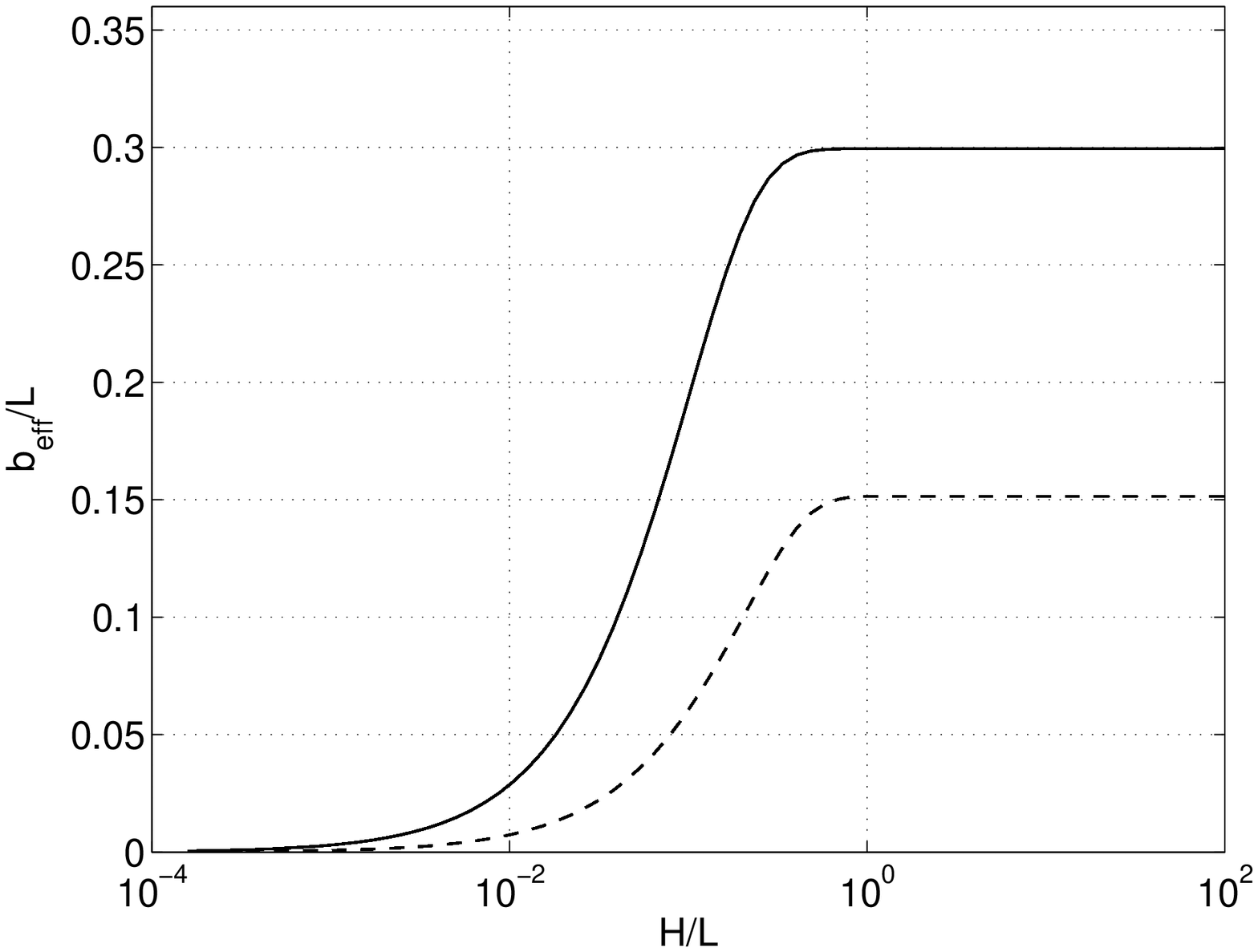}
(b) \includegraphics [width=0.45\textwidth]{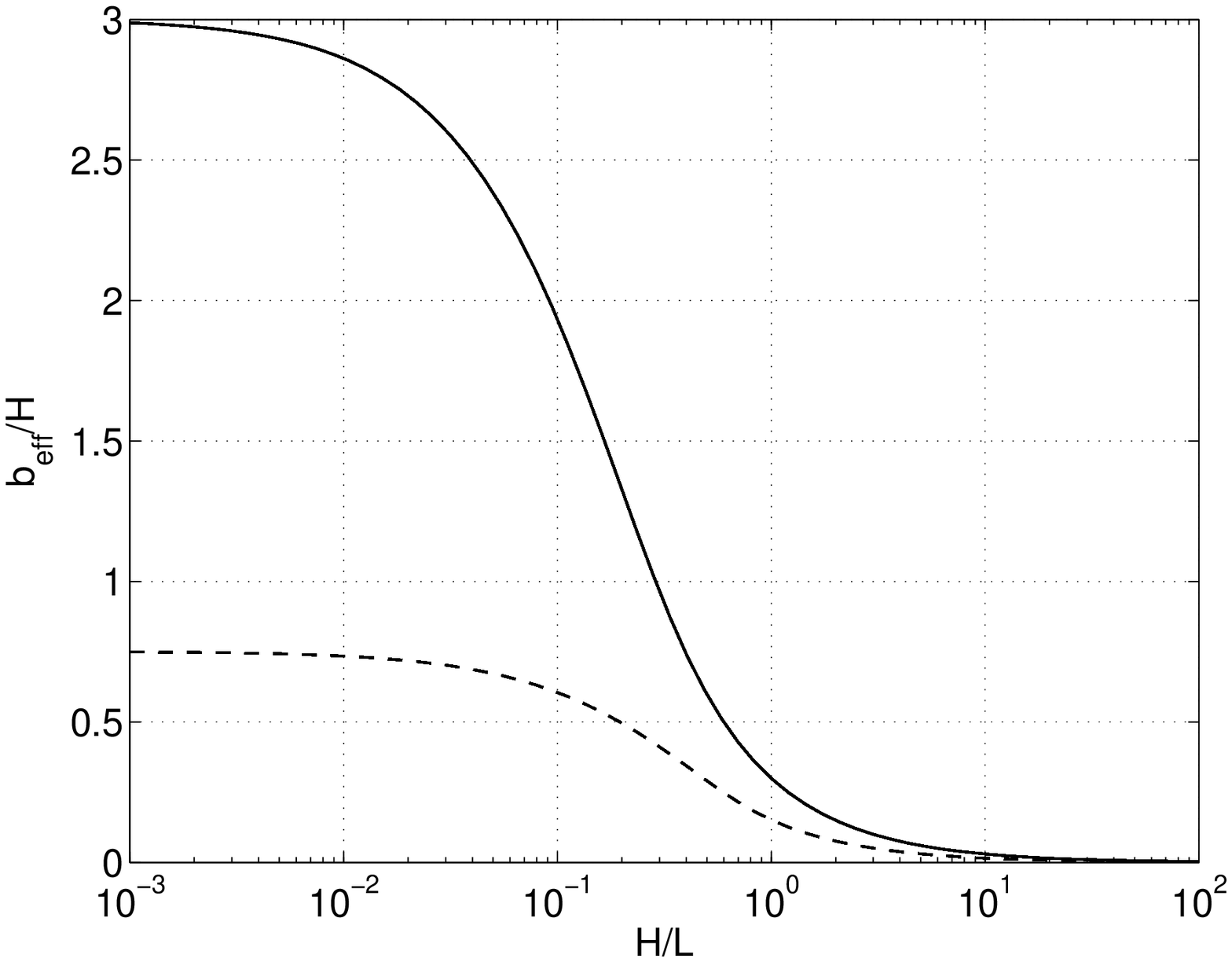}
\end{center}
\caption{
Eigenvalues $b^{\parallel}_{\rm eff}$ (solid curve) and $b^{\perp}_{\rm eff}$ (dashed curve) of the slip length tensor for stick-slip stripes of period $L$ with local slip length at liquid-gas interface $b/L=20$ and slipping area fraction $\phi_2=0.75$ as a function of the thickness of the channel, $H$. Reprinted with permission from~\cite{vinogradova.oi:2010}. Copyright (2011) by the IOP Publishing Ltd. }
  \label{fig:b_eff}
\end{figure}

The effective slip lengths in eigendirections (which are in this case obviously parallel and orthogonal to stripes) has been recently calculated~\cite{vinogradova.oi:2010} by using the dual series technique suggested in  work~\cite{sbragaglia.m:2007,belyaev.av:2010a}.
 Fig.~\ref{fig:b_eff}a shows the typical calculation results (the numerical example corresponds to $b/L=20$ and $\phi_2=0.75$), and demonstrates that the effective slip lengths increase with $H$ and saturate for a thick gap. This points to the fact that an effective boundary condition \emph{is not} a characteristic of liquid/solid interface solely, but depends on the flow configuration and interplay between typical length scales, $L$, $H$, and $b$, of the problem. Next we discuss asymptotic limits (of small and large gaps) of this semi-analytical solution.~\cite{vinogradova.oi:2010}

\subsubsection{Thin channel}

 In the limit of $H\ll L$ we get
\begin{equation}\label{beff_smallH}
   b_{\rm eff}^{\parallel} \simeq \frac{b H \phi_2}{H + b \phi_1},\quad b_{\rm eff}^{\perp} \simeq \frac{b H \phi_2}{H + 4b \phi_1}.
\end{equation}
These expressions are independent on $L$, but depend on $H$, and suggest to distinguish between two separate cases.

If $b\ll H$ we obtain
\begin{equation}\label{beff_smallH_limit1}
   b_{\rm eff}^{\perp} \simeq b_{\rm eff}^{\parallel}\simeq b\phi_2,
\end{equation}
so that despite the surface anisotropy we predict a simple surface averaged effective slip. Although this limit is less important for pressure-driven microfluidics, it may have
relevance for amplifying transport phenomena~\cite{ajdari.a:2006}.

When
$H\ll b$ we derive
\begin{equation}\label{beff_smallH_limit2}
  b_{\rm eff}^{\parallel} \simeq H \frac{\phi_1}{\phi_2},\quad b_{\rm eff}^{\perp} \simeq  \frac{1}{4} b_{\rm eff}^{\parallel}.
\end{equation}
The above formula implies the effective slip length is generally four times as large for parallel versus  perpendicular pressure driven flow. Both asymptotic results, Eqs.(\ref{beff_smallH_limit1}) and (\ref{beff_smallH_limit2}), are surprising taking into account that for anisotropic Stokes flow in a thick channel factor of two is often expected as reminiscent results for striped pipes~\cite{lauga.e:2003}, sinusoidal grooves~\cite{kamrin.k:2010} and the classical result that a rod sediments twice as fast in creeping flow if aligned vertically rather than horizontally~\cite{batchelor.gk:1970}. A very important conclusion from our analysis is that this standard scenario can significantly differ in a thin super-hydrophobic channel, by giving a whole spectrum of possibilities, from isotropic to highly anisotropic flow, depending on the ratio $b/H$.

\begin{figure}
\begin{center}
  (a)\includegraphics [width=7.5 cm]{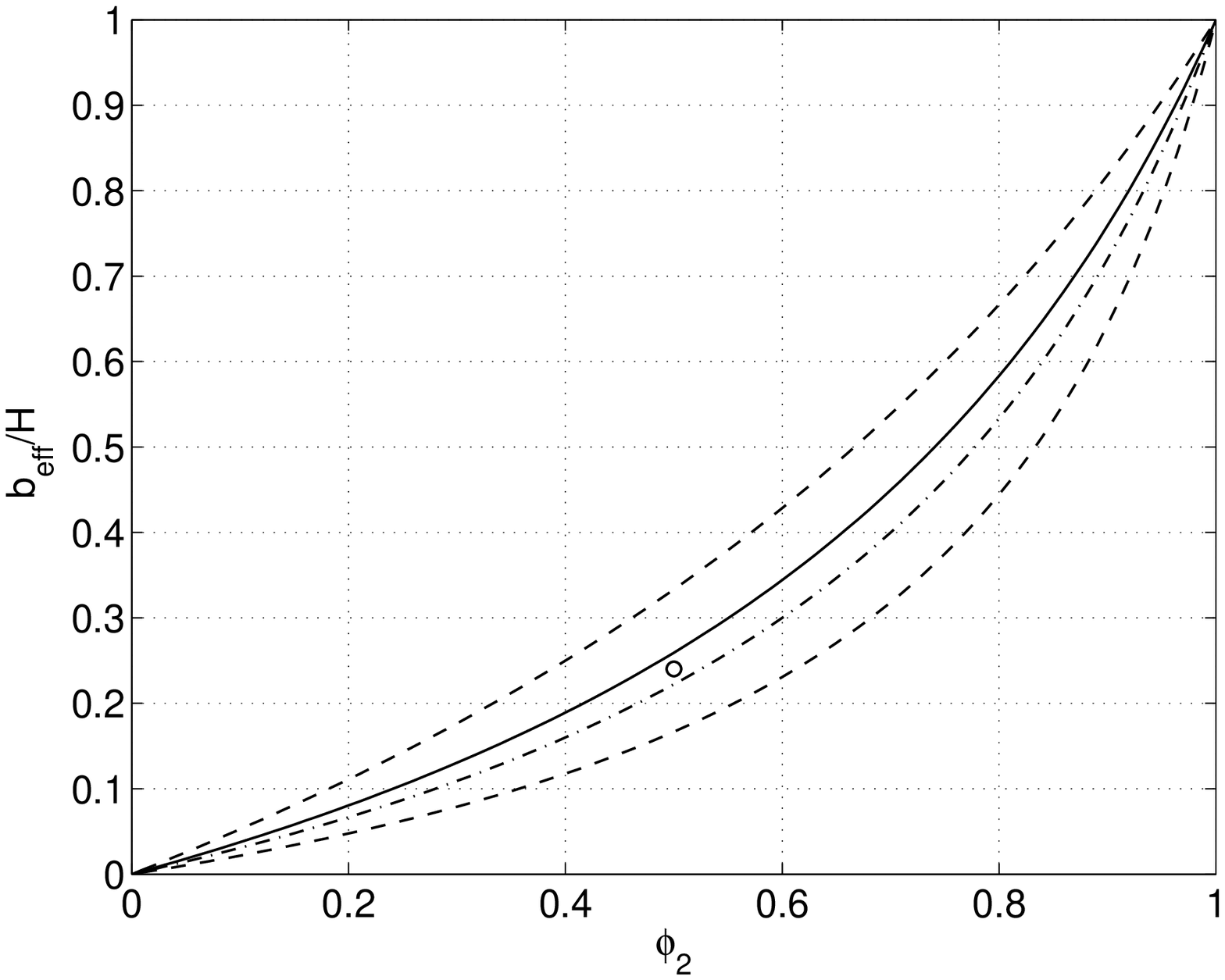}
  (b)\includegraphics [width=7.5 cm]{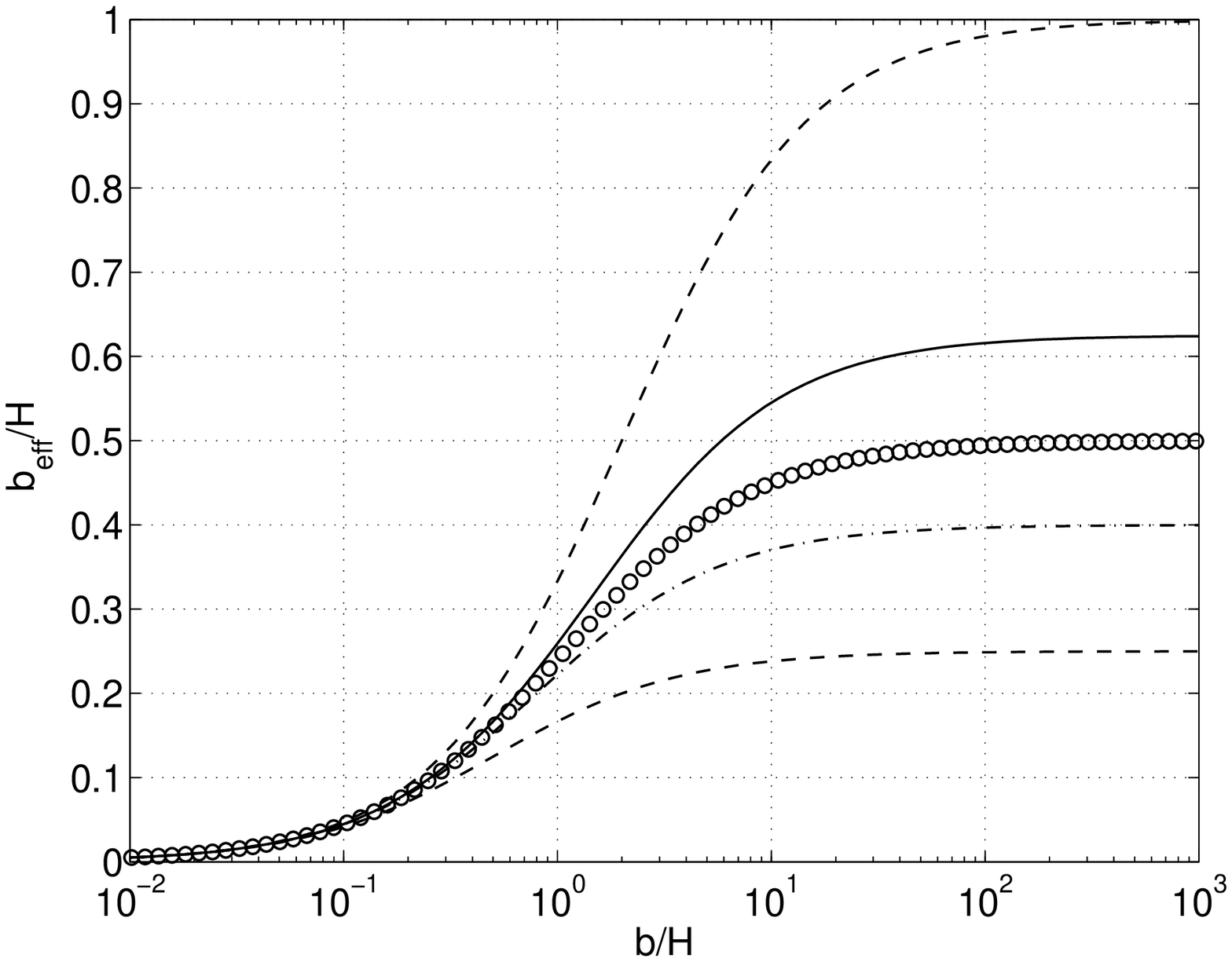}
  \end{center}
     \caption
     {Effective slip length, $b_{\rm eff}/H$, versus $\phi_2$ [for $b/H=1$] (a) and $b/H$ [for $\phi_2=0.5$] (b) in a thin gap limit, $H\ll L$. SH surfaces are: anisotropic stick-slip stripes attaining Wiener bounds (dashed curves), isotropic textures attaining Hashin-Shtrickman bounds (upper -- solid, lower -- dash-dotted curves) and satisfying the phase-interchange theorem (circles). Reprinted with permission from~\cite{vinogradova.oi:2010}. Copyright (2011) by the IOP Publishing Ltd.}
  \label{fig:b}
\end{figure}

Note that in case of a thin channel the flow can be described by an expression of Darcy's law, which relates the depth-averaged fluid velocity to an average pressure gradient along the plates through the effective permeability of the channel. The permeability, ${\bf K}_{\rm eff}$, is in turn expressed through effective slip length ${\bf b}_{\rm eff}$, and permeability and slip-length tensors are coaxial. Such an approach allows one to use the theory of transport in heterogeneous media~\cite{Torquato:2002}, which provides \emph{exact} results for an effective permeability over length scales much larger than the heterogeneity. This theory allows one to
derive rigorous bounds on an effective slip length for
arbitrary textures, given only the
area fraction and local (any) slip lengths of the low-slip ($b_1$) and high-slip ($b_2$)
regions~\cite{feuillebois.f:2009,feuillebois.f:2010}. These bounds constrain the attainable effective slip and provide theoretical guidance for texture optimization, since they are attained only by certain special textures in the theory. In some regimes, the bounds are close enough to obviate the need for tedious
calculations of flows over particular textures. In particular, by using the general result of the theory of bounds~\cite{feuillebois.f:2009} one can easily derive Eq.(\ref{beff_smallH}) and its limits, as well as to prove that
for a thin channel longitudinal (transverse) stripes satisfy upper (lower) Wiener bounds, i.e.
provide the largest (smallest) possible slip that can be
achieved by any texture. We remark and stress that according to results~\cite{feuillebois.f:2009} stripes should not be necessarily periodic.

Typical dependence of Wiener bounds for $b_{\rm eff}/H$ on $\phi_2$ (at fixed $b/H$) and on $b/H$ (at fixed $\phi_2$) is shown in Fig.~\ref{fig:b}, which well illustrates that the key parameters determining
effective slip in the thin channel is the area fraction of solid, $\phi_1$, in contact
with the liquid. If this is very small (or $\phi_2\to 1$), for
all textures the effective slip tends to a maximum value, $b_{\rm eff} \to b$. We can also conclude that maximizing
$b$ also plays a important role to achieve large effective slip.

\subsubsection{Thick channel}

In the opposite case of infinitely large thickness ($H \gg L$) the effective slip lengths are~\cite{belyaev.av:2010a}
\begin{equation}\label{beff_par_largeH}
  b_{\rm eff}^{\parallel} \simeq \frac{L}{\pi} \frac{\ln\left[\sec\left(\displaystyle\frac{\pi \phi_2}{2 }\right)\right]}{1+\displaystyle\frac{L}{\pi b}\ln\left[\sec\displaystyle\left(\frac{\pi \phi_2}{2 }\right)+\tan\displaystyle\left(\frac{\pi \phi_2}{2}\right)\right]},
\end{equation}
\begin{equation}\label{beff_ort_largeH}
  b_{\rm eff}^{\perp} \simeq \frac{L}{2 \pi} \frac{\ln\left[\sec\left(\displaystyle\frac{\pi \phi_2}{2 }\right)\right]}{1+\displaystyle\frac{L}{2 \pi b}\ln\left[\sec\displaystyle\left(\frac{\pi \phi_2}{2 }\right)+\tan\displaystyle\left(\frac{\pi \phi_2}{2}\right)\right]}.
\end{equation}
The above results apply for a single surface, and are independent on $H$. However, these expressions for effective slip lengths depend strongly on a texture period $L$. When $b \ll L$ we again derives the area-averaged slip length, Eq.(\ref{beff_smallH_limit1}). When $b \gg L$,  expressions (\ref{beff_par_largeH}) and (\ref{beff_ort_largeH}) take form
\begin{equation}\label{beff_ort_largeH_id}
  b_{\rm eff}^{\perp} \simeq \frac{L}{2 \pi} \ln\left[\sec\left(\displaystyle\frac{\pi \phi_2}{2 }\right)\right],\quad b_{\rm eff}^{\parallel}\simeq2 b_{\rm eff}^{\perp},
\end{equation}
that coincides with the result obtained by Lauga and Stone~\cite{lauga.e:2003} for the ideal slip ($b\to \infty$) case. We stress that the commonly expected factor of two for the ratio of $b_{\rm eff}^{\parallel}$ and $b_{\rm eff}^{\perp}$ is predicted only for very large $b/L$. In all other situation the anisotropy of the flow is smaller and even disappears at moderate $b/L$.

\subsection{Isotropic surfaces.}

As stressed above most solids are isotropic, i.e. without a preferred direction. Unfortunately, from the hydrodynamic point of view, this situation is more complicated than considered above. Below we discuss only some aspects of the hydrodynamic behavior in the thin and thick channel situations. For a thin channel, arguments  are based on the already mentioned theory of transport in heterogeneous media~\cite{Torquato:2002} and derived bounds on an effective slip length   (effective slip length and permeability tensors are now becoming simply proportional to the unit tensor ${\bf I}$) for arbitrary isotropic textures~\cite{feuillebois.f:2009,feuillebois.f:2010}. For a thick channel, the only available arguments are based on the scaling theory and numerical calculations~\cite{ybert.c:2007}, which however provide us with some guidance.

\subsubsection{Thin channel}

If the only knowledge about the two-phase isotropic texture is $\phi_1$, $\phi_2$,
then the Hashin-Shtrikman (HS) bounds apply for the
effective permeability, by giving the corresponding upper and lower  HS bounds for the effective slip length~\cite{feuillebois.f:2009,feuillebois.f:2010}. These bounds can be attained by the
special HS fractal pattern sketched in Fig.~\ref{fig:designs}b. For one bound, space is filled by disks of all sizes, each containing
a circular core of one component and a thick ring of the
other (with proportions set by the concentration), and switching
the components gives the other bound. Fractal geometry is not
necessary, however, since periodic honeycomb-like structures
can also attain the bounds. The general solution~\cite{feuillebois.f:2009,feuillebois.f:2010} allows to deduce a consequential analytical results for an asymmetric case considered in this paper. The upper (HS) bound can be then presented as
\begin{equation}\label{b_HSU}
    b_{\rm eff} = \displaystyle \frac{b H \phi_2 (2 H + 5 b)}{H (2 H + 5 b) + b \phi_1 (5 H + b)},
    \end{equation}
and the lower (HS) bound reads
    \begin{equation}\label{b_HSL}
b_{\rm eff} = \frac{2 b H \phi_2}{2 H + 5 b \phi_1}.
\end{equation}
At small $b/H$ we get Eq.(\ref{beff_smallH_limit1}), and at large $b/H$ these give for upper and lower bounds
\begin{equation}\label{b_HS_largeb}
    b_{\rm eff} = \displaystyle \frac{5 H \phi_2}{8 \phi_1}, \quad {\rm and} \quad b_{\rm eff} = \displaystyle \frac{2 H \phi_2}{5 \phi_1},
    \end{equation}
correspondingly.

Finally, phase interchange results~\cite{feuillebois.f:2009} can be applied for some specific patterns (Fig.~\ref{fig:designs}c,d). The phase interchange theorem states that the effective permeability ${\bf K}_{\rm eff}(b_1,b_2)$ of the medium, when rotated by $\pi/2$, is related to the effective permeability of the medium obtained by interchanging phases 1 and 2, viz. ${\bf K}_{\rm eff}(b_2,b_1)$:
\[
 [ {\bf R} \cdot {\bf K}_{\rm eff}(b_1, b_2) \cdot {\bf R}^t ]
\cdot
{\bf K}_{\rm eff}(b_2, b_1) =  K_1 K_2 {\bf I}
\]
where $b_{1,2}$ are the local slip lengths for each medium, ${\bf R}$ is the rotation tensor and ${\bf R}^t$ is its transpose. In the particular case of a medium which is invariant by a $\pi/2$ rotation followed by a phase interchange, the classical result follows:
\[
 K_{\rm eff} = \sqrt{K_1 K_2}
\]
Obviously, $\phi_1=\phi_2=0.5$ for such a medium so that:
\begin{equation}
  b_{\rm eff}= \frac{3 H}{\displaystyle 4-\sqrt{1+\frac{3 b}{H+b}}} -H,
\label{phase_interchange}
\end{equation}
At $b/H \ll 1$ we again derive Eq.(\ref{beff_smallH_limit1}), indicating that at this limit all textures show a kind of universal behavior and the effective slip coincides with the average. This suggests that the effective slip is controlled by the smallest scale of the problem~\cite{bocquet2007,belyaev.av:2010a}, so that at this limit $b_{\rm eff}$ is no longer dependent on $H$, being proportional to $b$ only.
If $b/H \gg 1$ we simply get
\begin{equation}
  b_{\rm eff}= \frac{ H}{2}
\end{equation}
again suggesting a kind of universality, i.e. similarly to anisotropic stripes (cf. Eq.~\ref{beff_smallH_limit2}), in this limit $b_{\rm eff}/H$ for all isotropic textures almost likely scale as $\propto \phi_2/\phi_1$.

The results for these special textures are included in Fig.~\ref{fig:b}, which shows that Hashin-Strickman bounds are relatively
close and confined between Wiener ones. For completeness, we give in Table~\ref{tbl:b_eff_thin} a summary of main expressions for an effective slip in a thin channel.

\begin{table}[h]
  \tbl{The effective slip length $b_{\rm eff}$ for different textures (shown in Fig.~\ref{fig:designs}) in a thin gap limit ($H\ll L$)}
  {\begin{tabular}{@{}lc@{}}
  \toprule
  Texture &  $b_{\rm eff}$  \\
  \colrule
  Wiener upper bound (longitudinal stripes) & $\displaystyle\frac{b H \phi_2}{H + b \phi_1}$ \\
  Wiener lower bound (transverse stripes) & $\displaystyle\frac{b H \phi_2}{H + 4b \phi_1}$ \\
  Hashin-Shtrickman upper bound (Hashin-Shtrickman fractal, honeycomb-like texture) & $\displaystyle \frac{b H \phi_2 (2 H + 5 b)}{H (2 H + 5 b) + b \phi_1 (5 H + b)}$\\
  Hashin-Shtrickman lower bound (Hashin-Shtrickman fractal, honeycomb-like texture) & $\displaystyle \frac{2 b H \phi_2}{2 H + 5 b \phi_1}$\\
  Phase interchange patterns (Schulgasser texture, family of chessboards) & $\displaystyle\frac{3 H}{\displaystyle 4-\sqrt{1+3 b/(H+b)}} -H$  \\
  \botrule
  \end{tabular}}\label{tbl:b_eff_thin}
\end{table}

\subsubsection{Thick channel}

For this situation the exact solution was not found so far. Nevertheless, some simple scaling
expressions have been proposed for a geometry of
pillars~\cite{bocquet2007,ybert.c:2007}, by predicting $b_{\rm eff} \propto L / (\pi \sqrt{\phi_1})$ (cf. scaling results for stripes $b_{\rm eff} \propto L / \ln (1/\phi_1)$). This simple result would deserve some analytical justification, which has not been performed up to now, despite some recent approximate analysis.~\cite{davis.amj:2010} Recent semi-analytical and numerical results confirm this scaling dependence,~\cite{ng.co:2010b} which, in particular, suggests that in a thick channel the array of pillars  will give larger effective slip than longitudinal stripes for
a sufficiently small solid fraction. This is opposite to a prediction for a thin channel, where an array of longitudinal stripes provides the
largest possible slip that can be achieved by any texture,
whether isotropic or anisotropic.

\section{Other special properties of super-hydrophobic surfaces}

As we see above, hydrophobic Cassie materials generate large and anisotropic effective slippage compared to simple, smooth channels, ideal situation for various potential applications. A straightforward implication of super-hydrophobic slip would be the great reduction of the viscous drag of thin microchannels (enhanced \emph{forward} flow), and some useful examples can be found in~\cite{Bazant08}. Below we illustrate the potential of super-hydrophobic surfaces and possibilities of the effective slip approach by discussing a couple of other applications. Namely, we show that optimized super-hydrophobic textures may be successfully used in a passive microfluidic mixing and for a reduction of a hydrodynamic drag force.

\subsection{Transverse flow}

The effective hydrodynamic slip~\cite{stone2004,Bazant08,kamrin.k:2010} of anisotropic textured surfaces is generally tensorial, which is due to secondary flows {\it transverse} to the direction of the applied pressure gradient. In the case of grooved no-slip surfaces (Wenzel state), such a flow has been analyzed for small height variations~\cite{stroock2002a} and thick channels~\cite{wang2003}, and herringbone patterns have been designed to achieve passive chaotic mixing during pressure-driven flow through a microchannel~\cite{stroock2002b,stroock2004}.

\begin{figure}
\begin{center}
(a)\includegraphics[width=9 cm,clip]{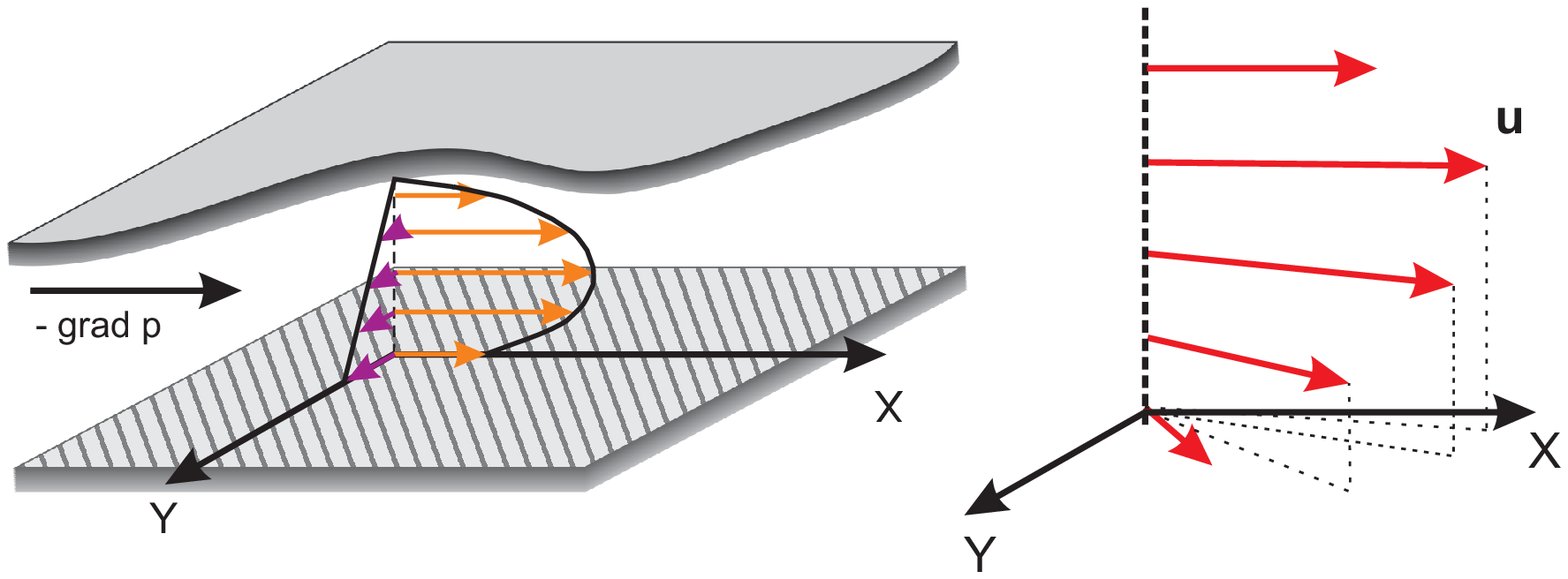}
(b)\includegraphics[width=9 cm,clip]{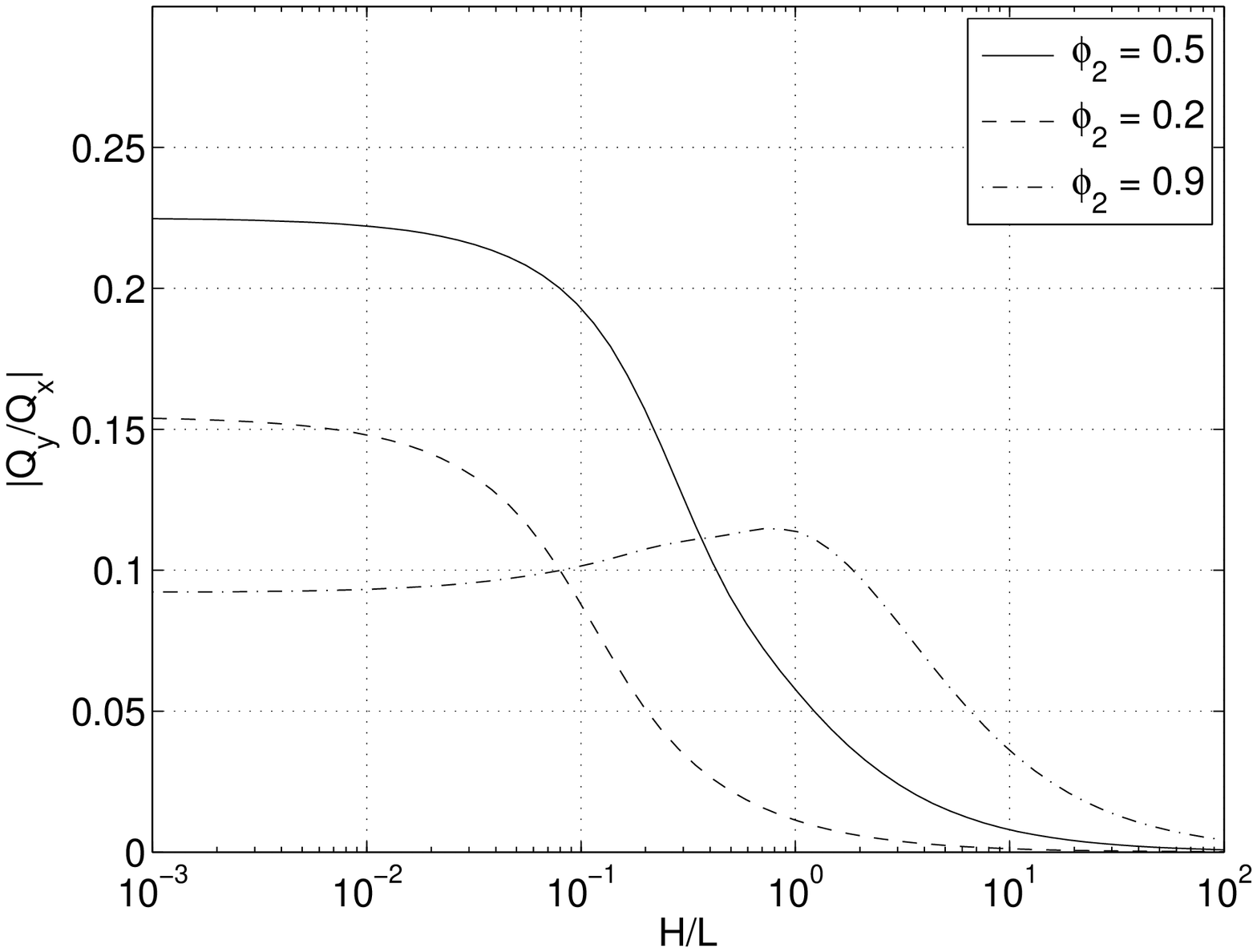}
\end{center}
\caption{(a) Scheme of a transverse flow generation. (b) Fraction of flow vector $\textbf{Q}$ components (maximized over $\theta$) as function of channel thickness for local slip $b/L=1000$ and slip fraction $\phi_2=0.5$ (solid line), $0.2$ (dashed) and $0.9$ (dash-dot). Reprinted with permission from~\cite{vinogradova.oi:2010}. Copyright (2011) by the IOP Publishing Ltd.}
\label{fig:transvflow}
\end{figure}

It has recently been demonstrated that similar effects may be generated by a super-hydrophobic Cassie surface. The transverse flow due to surface anisotropy is generated only in the vicinity of the wall and disappears far from it, which has already been observed in experiment~\cite{ou.j:2007}. In other words, the effective velocity profile is `twisted'  close to the super-hydrophobic wall (see Fig.~\ref{fig:transvflow}a).

To evaluate the transverse flow the velocity profile has been integrated across the channel to obtain the flow vector:
\begin{equation}\label{Q}
    \textbf{Q}=\int\limits_0^H{\left\langle\textbf{u}(z)\right\rangle dz}
\end{equation}
with the components:
\begin{equation}\label{Q_X_I}
    Q_x=\frac{\sigma}{\eta} \frac{H^3}{12} \left[1+3 \frac{ \left(H b_{\rm eff}^{\parallel} \cos^2{\theta} + H b_{\rm eff}^{\perp} \sin^2{\theta} + b_{\rm eff}^{\parallel} b_{\rm eff}^{\perp}\right)}{(H+b_{\rm eff}^{\parallel})(H+b_{\rm eff}^{\perp})} \right],
\end{equation}
\begin{equation}\label{Q_Y_I}
    Q_y=\frac{\sigma}{\eta} \frac{H^4}{4} \frac{(b_{\rm eff}^{\parallel}-b_{\rm eff}^{\perp})\sin{\theta}\cos{\theta}}{(H+b_{\rm eff}^{\parallel})(H+b_{\rm eff}^{\perp})}.
\end{equation}
To optimize the texture, channel thickness and the angle  $\theta$ between the directions of stripes and the pressure gradient, so that $|Q_y/Q_x|$ is maximum providing the  best transverse flow.

The maximization in respect to $\theta$ indicates that the optimal angle is,~\cite{vinogradova.oi:2010}
\begin{equation}\label{theta_max}
    \theta_{\rm max}=\pm \arctan \left[\frac{(1+4 b_{\rm eff}^{\parallel} /H)(1+b_{\rm eff}^{\perp}/H)}{(1+b_{\rm eff}^{\parallel} / H)(1+4 b_{\rm eff}^{\perp} / H)}\right]^{1/2}.
\end{equation}
and the value of the maximum reads
\begin{equation}\label{theta_max2}
    \left|\frac{Q_y}{Q_x}\right|= \frac{1}{2}\left(\tan\theta_{\rm max}-\frac{1}{\tan\theta_{\rm max}}\right).
\end{equation}
We conclude, therefore, that since $H$ is fixed, the maximal $|Q_y/Q_x|$ corresponds to the largest physically possible $b$, i.e. the perfect slip at the gas sectors.

To optimize the fraction   of the slipping area, $\phi_2$, we should now exploit results for effective slip lengths $b_{\rm eff}^{\parallel,\perp}$ obtained above. Fig.\ref{fig:transvflow}b shows the computed value of $|Q_y/Q_x|$ \emph{vs.}
 $H/L$ for several $\phi_2$. The calculations are made using the value of $\theta$ defined by Eq.(\ref{theta_max}). The data suggest that the effect of $\phi_2$ on a transverse flow depends on the thickness of the channel. For a thick gap the increase in gas fraction, $\phi_2$, augments a transverse flow. This result has a simple explanation. For an infinite channel $b_{\rm eff}^{\parallel, \perp}/H \ll 1$ (see Fig.(\ref{fig:b}b)), which gives
\begin{equation}\label{Qfrac1}
    \left| \frac{Q_y}{Q_x} \right|_{H\rightarrow \infty} \simeq \frac{3}{2}\frac{\Delta b_{\rm eff}}{H},
\end{equation}
i.e. in a thick channel the amplitude of a transverse flow is controlled by the difference between eigenvalues of the effective slip tensor, $\Delta b_{\rm eff}= b_{\rm eff}^{\parallel}-b_{\rm eff}^{\perp}$, which increases with $\phi_2$ as follows from the above analysis. We stress however, that since $|Q_y/Q_x| \propto H^{-1}$, the mixing in a thick super-hydrophobic channel would be not very efficient. A more appropriate situation corresponds to a thin channel as it is well illustrated in Fig.\ref{fig:transvflow}b. We see, that the largest transverse flow can be generated at intermediate values of $\phi_2$. The limit of thin channel has recently been studied in details by using a general theory of mathematical bounds~\cite{Torquato:2002}, and the optimum value of $\phi_2=0.5$ corresponding a numerical example in Fig.\ref{fig:transvflow}b has been rigorously derived~\cite{mixer2010}.

An important conclusion from the above analysis is that the surface textures which optimize transverse flow can significantly differ from those optimizing effective (forward) slip. It is well known, and we additionally demonstrated above, that the effective slip of a super-hydrophobic surface is maximized by reducing the solid-liquid area fraction $\phi_1$. In contrast, we have shown that transverse flow in super-hydrophobic channels is maximized by stripes with a rather large solid fraction, $\phi_1=0.5$, where  the effective slip is relatively small.

\subsection{Hydrodynamic interactions}

\begin{figure}
\begin{center}
  \includegraphics [width=5.5 cm]{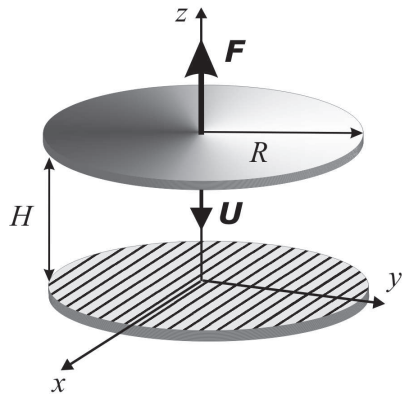}\,\,\,
    \end{center}
  \caption{Sketch of a hydrophilic disk approaching super-hydrophobic striped disk. }
  \label{fig:disks}
\end{figure}

As a consequence of the superlubrication potential, a hydrophobic texture could significantly modify squeeze film drainage between surfaces. It is of obvious practical interest to minimize the hydrodynamic resistance, $F$, to approach of surfaces.

For a Reynolds problem, where a disk of radius $R$ is driven towards (in our case) a super-hydrophobic plane with a velocity $U$ (see Fig.~\ref{fig:disks}) we should solve Eqs.(\ref{Stokes}) by applying the effective tensorial boundary condition, Eq.(\ref{effBC}), at the super-hydrophobic wall. This allows to derive a general expression for hydrodynamic force satisfying the condition $p=p_0$ at the edge of the disk~\cite{belyaev.av:2010b}
\begin{equation}\label{force1}
  F= \frac{3}{2}\frac{\pi\eta U R^4}{H^3} f^{\ast}_{\rm eff}=F_R f^{\ast}_{\rm eff},
\end{equation}
where $F_R$ represents the classical solution of creeping flow equations of the Reynolds lubrication theory~\cite{reynolds1886}, and the correction for an effective slip is
\begin{equation}\label{force2}
  f^{\ast}_{\rm eff}=\frac{F}{F_R}= 2\left[\frac{H+4 b_{\rm eff}^{\parallel}(H)}{H+b_{\rm eff}^{\parallel}(H)}+\frac{H+4 b_{\rm eff}^{\perp}(H)}{H+ b_{\rm eff}^{\perp}(H)}\right]^{-1}.
\end{equation}
Thus the effective correction for a super-hydrophobic slip is the harmonic mean
of corrections expressed through effective slip lengths in two principal directions,
\begin{equation}\label{fast_sum}
    f^{\ast}_{\rm eff}=2 \left(\frac{1}{f^{\ast,\parallel}_{\rm eff}}+\frac{1}{f^{\ast,\perp}_{\rm eff}}\right)^{-1}
\end{equation}

In case of isotropic textures, all directions are equivalent with $b_{\rm eff}^{\parallel}=b_{\rm eff}^{\perp}=b_{\rm eff}$, so we get
\begin{equation}\label{force3}
 f^{\ast}_{\rm eff}= \frac{F}{F_R}= \frac{H+b_{\rm eff}(H)}{H+4 b_{\rm eff}(H)}
\end{equation}
Obviously, the case $b_{\rm eff}^{\parallel}=b_{\rm eff}^{\perp}=0$ corresponds to $f^{\ast}_{\rm eff}=1$ and gives the Reynolds formula.

    The expression for $f^{\ast}_{\rm eff}$ is very general and relates it to the effective slip length of the super-hydrophobic wall and the gap. In order to quantify the reduction of a drag force due to a presence of a super-hydrophobic wall, this expression can be used for \emph{all} anisotropic and isotropic textures, where analytical or numerical expressions for $b_{\rm eff}^{\parallel, \perp}$ have been obtained.

\begin{figure}
  \begin{center}
  (a)\includegraphics [width=7 cm]{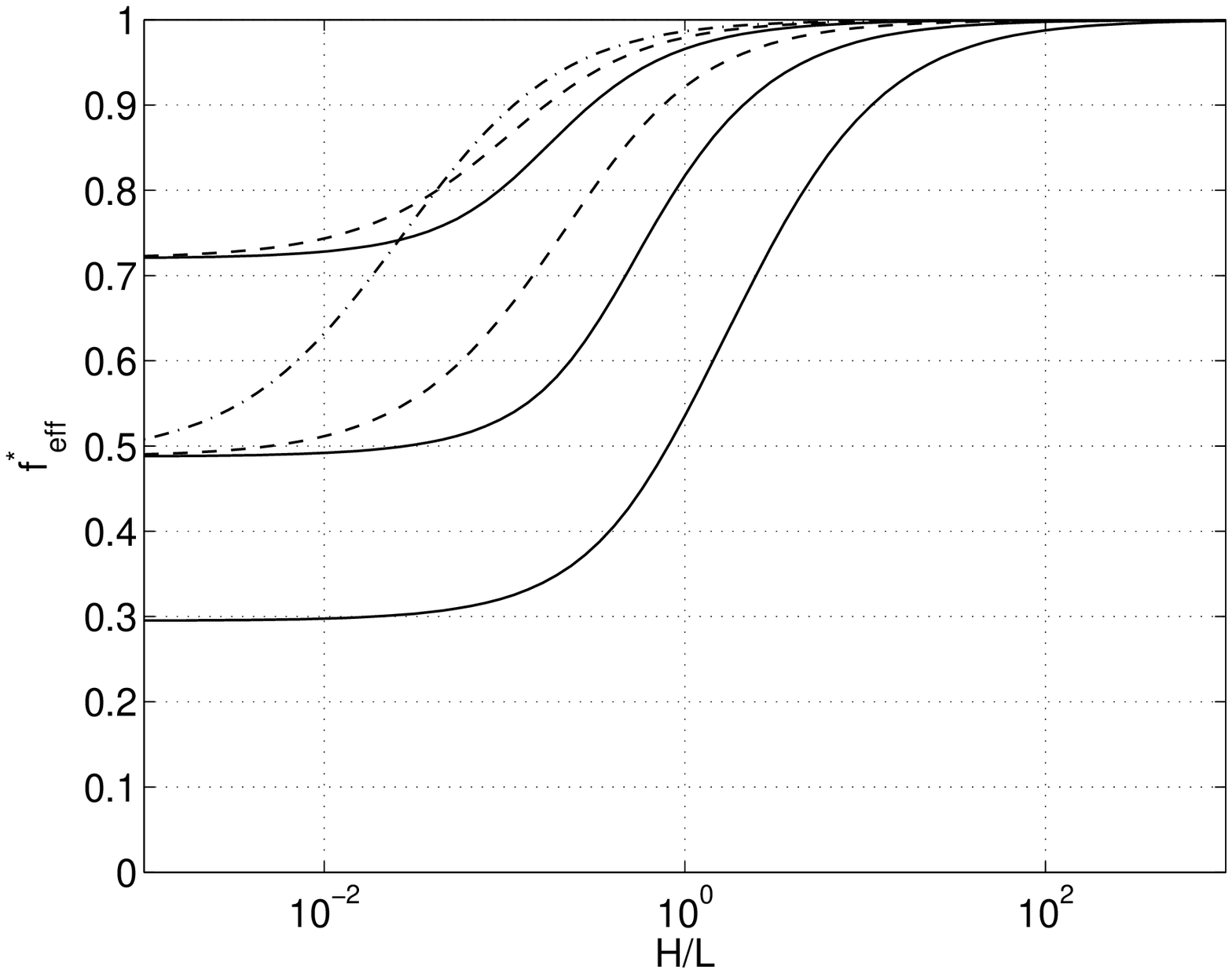}
  (b)\includegraphics [width=7 cm]{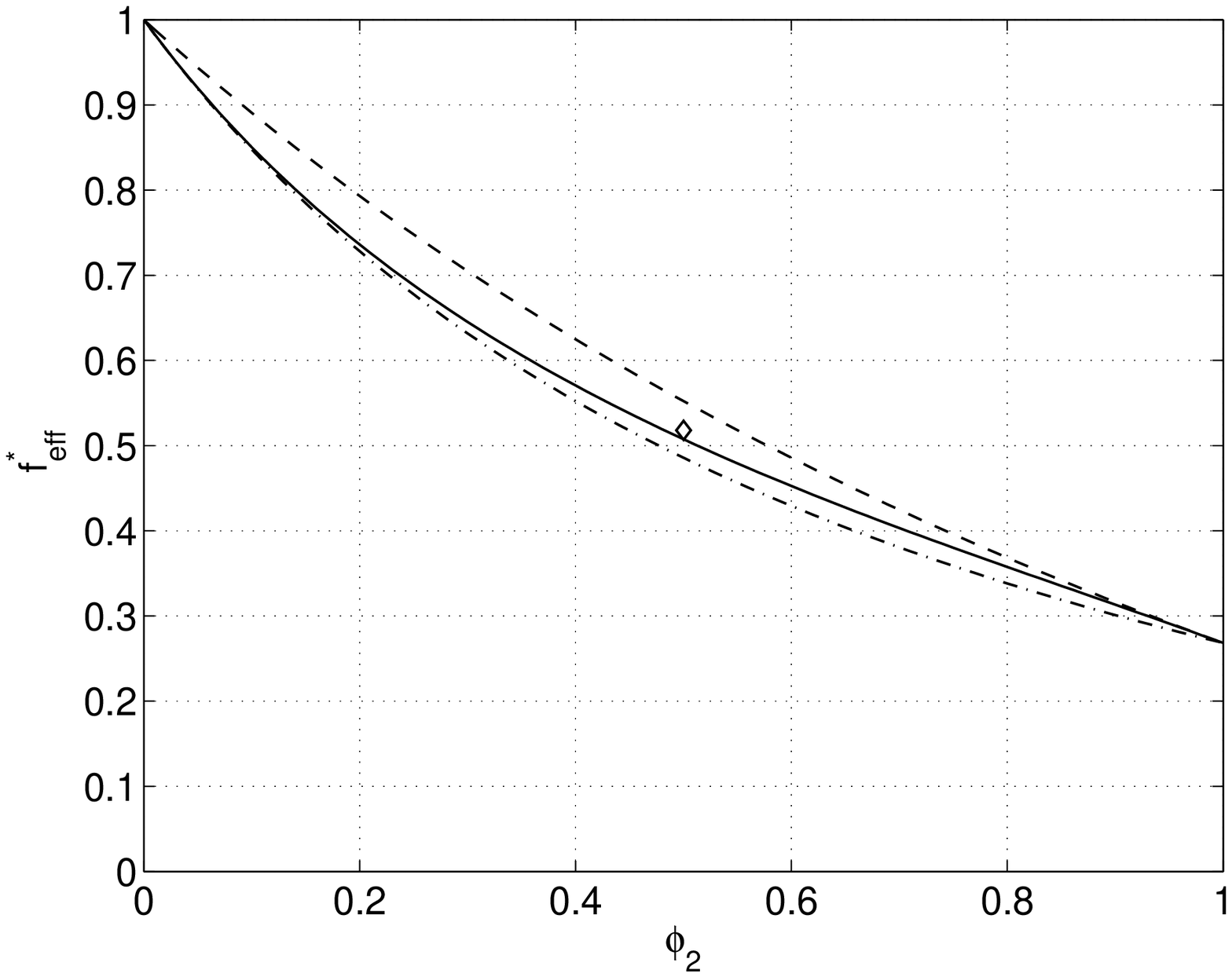}
  \end{center}
  \caption{(a) The correction factor $f^{\ast}_{\rm eff}=F/F_R$ for hydrodynamic resistance force exerted on disk interacting with super-hydrophobic stripes vs. dimensionless gap width $H/L$. Solid curves correspond to local slip length $b/L=10$ (from top to bottom $\phi_2=0.2$, $0.5$ and $0.9$), dashed curves -- to $b/L=0.1$ (from top to bottom $\phi_2=0.2$ and $0.5$), dash-dotted curve - to $b/L=0.01$ and $\phi_2=0.5$. (b) The plot of $f^{\ast}_{\rm eff}$ versus $\phi_2$ for a thin gap ($H\ll L$) and several super-hydrophobic patterns: anisotropic stripes (solid line), isotropic textures attaining Hashin-Strickman bounds (dashed and dash-dotted lines) and isotropic Schulgasser structure (diamond), all with local slip $b/H = 10$. Reprinted with permission from~\cite{vinogradova.oi:2010}. Copyright (2011) by the IOP Publishing Ltd.  }
  \label{fig:disks1}
\end{figure}

An important consequence of Eq.(\ref{force2}) is that to reduce a drag force we need to maximize the ratio $b_{\rm eff}/H$, but not the absolute values of effective slip itself.
This is illustrated in Fig.~\ref{fig:disks1}a, where values presented in Fig.~\ref{fig:b_eff}b were used to compute the correction
for effective slip, $f^{\ast}_{\rm eff}$ as a function of the gap. At small $H/L$ our calculations reproduce the asymptotic values predicted by Eqs.~(\ref{beff_smallH_limit2}). They however vanish at large distances, where $b_{\rm eff}^{\parallel, \perp}/H$ are getting negligibly small. The useful analytical expressions for $f^{\ast}_{\rm eff}$ corresponding to a configuration of stripes are presented in Table~\ref{tbl:asympt}. Similar estimates for the most important situation of a thin gap can similarly be done for some isotropic textures, and we include these results into Table~\ref{tbl:isotr}.

The results presented in Tables~\ref{tbl:asympt} and \ref{tbl:isotr} show that the key parameter determining
reduction of drag is the area fraction of gas, $\phi_2$, in contact with
the liquid. This is illustrated in Fig.~\ref{fig:disks1}b, where (using a relatively
large $b/H$) Hashin-Strickman bounds for $f^{\ast}_{\rm eff}$
are plotted versus
$\phi_2$. If this is very small (or $\phi_1 \to 1$) for
all textures, the correction for slip tends to its absolute
maximum, $f^{\ast}_{\rm eff} \to 1$. In the most interesting limit, $\phi_2 \to 1$, we can
achieve the minimum possible value of correction for effective
slip, $f^{\ast}_{\rm eff} \to 1/4$
provided $b/H$ is large enough. We also stress that
the results for stripes are confined between Hashin-Strickman bounds for $f^{\ast}_{\rm eff}$. In other words, isotropic textures might be the best candidates for a reduction of a drag force.

Another important point to note would be that there exists no universal relationship between $f^{\ast}_{\rm eff}$ and the effective Cassie angle, $\cos\theta^*_{C}$ , of the super-hydrophobic surface. Indeed, according to the Cassie equation, $\cos\theta^*_{C}= (1 - \phi_{2}) \cos\theta -  \phi_{2},$~\cite{quere.d:2005} where $ \cos\theta$ is the thermodynamic contact angle at the smooth surface. It is well seen already from Fig.~\ref{fig:disks1}b that the dependence of $f^{\ast}_{\rm eff}$ on $\phi_{2}$ is much more complex. In particular, at fixed $H$ it depends on texture. In general, this also reflects the fact that there is no universal relationship between $b_{\rm eff}$ and $\phi_2$, despite some recent attempts to establish it.~\cite{voronov.rs:2008}.

\begin{table}[h]
  \tbl{Asymptotic expansions for the force correction factor $f^{\ast}_{\rm eff}$ in case of a striped surface.}
  {\begin{tabular}{@{}ll@{}}
  \toprule
  Limiting case &  $f^{\ast}_{\rm eff}$  \\
  \colrule
  $H \gg {\rm max}\{L, b\}$ & $1- \displaystyle\frac{3(b_{\rm eff}^{\parallel}+b_{\rm eff}^{\perp})}{2H}$ \\
  $L\ll H\ll b$ & $\displaystyle\frac{1}{4} +  \displaystyle\frac{9}{32} \frac{\pi H}{L \,\ln(\sec(\pi\phi_2/2))}$\\
  $b \ll H \ll L$ &  $\displaystyle1- \frac{3 b\phi_2}{H} $ \\
  $H \ll {\rm min}\{L, b\}$ & $\displaystyle\frac{2(4-3\phi_2)}{8+9\phi_2-9\phi_2^2}$  \\
  \botrule
  \end{tabular}}\label{tbl:asympt}
\end{table}

\begin{table}[h]
  \tbl{Correction factor $f^{\ast}_{\rm eff}$ for some specific isotropic patterns in a thin gap limit.}
  {\begin{tabular}{@{}lcc@{}}
  \toprule
  Pattern  & $b \ll H \ll L$  &  $H \ll {\rm min}\{L, b\}$   \\
  \colrule
  Hashin-Strickman upper bound  & $\displaystyle1- \frac{3 b\phi_2}{H}$  &  $\displaystyle\frac{5-3\phi_2}{5+3\phi_2}$  \\
  Hashin-Strickman lower bound &  $\displaystyle1- \frac{3 b\phi_2}{H} $ & $\displaystyle\frac{8-3\phi_2}{4(2+3\phi_2)}$ \\
  Phase interchange textures & $\displaystyle1- \frac{3 b}{2H}$  & $\displaystyle\frac{1}{2}$  \\
  \botrule
  \end{tabular}}\label{tbl:isotr}
\end{table}

\section{Interfacial transport phenomena}

\subsection{Hydrophobic surface}

Besides this drag-reduction potential in pressure-driven
flows, it has recently been predicted that hydrophobic slippage could be
best exploited in surface-driven transport.

\begin{figure}
\begin{center}
  (a) \includegraphics [width=6.5 cm]{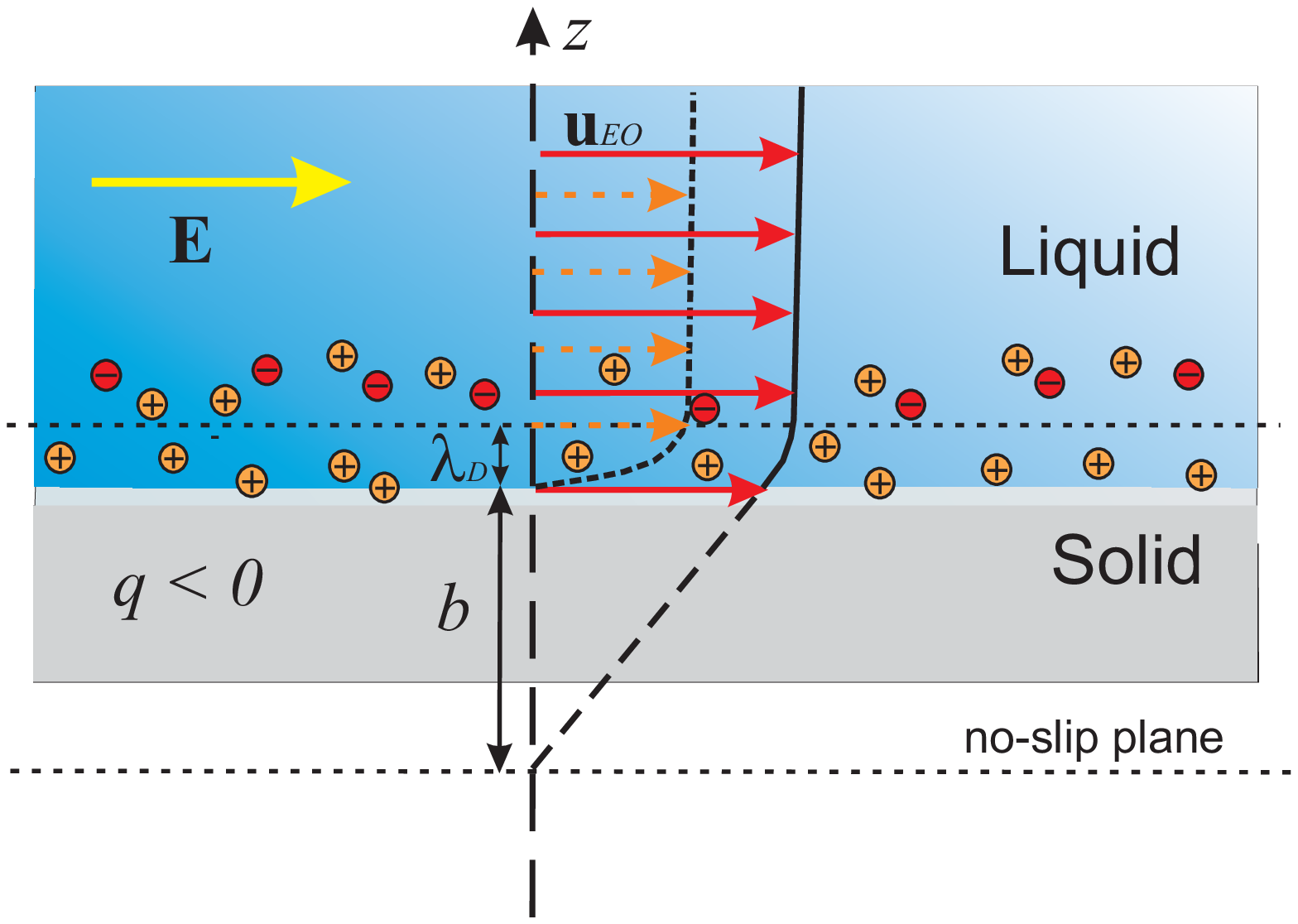}\\
  (b) \includegraphics [width=6.5 cm]{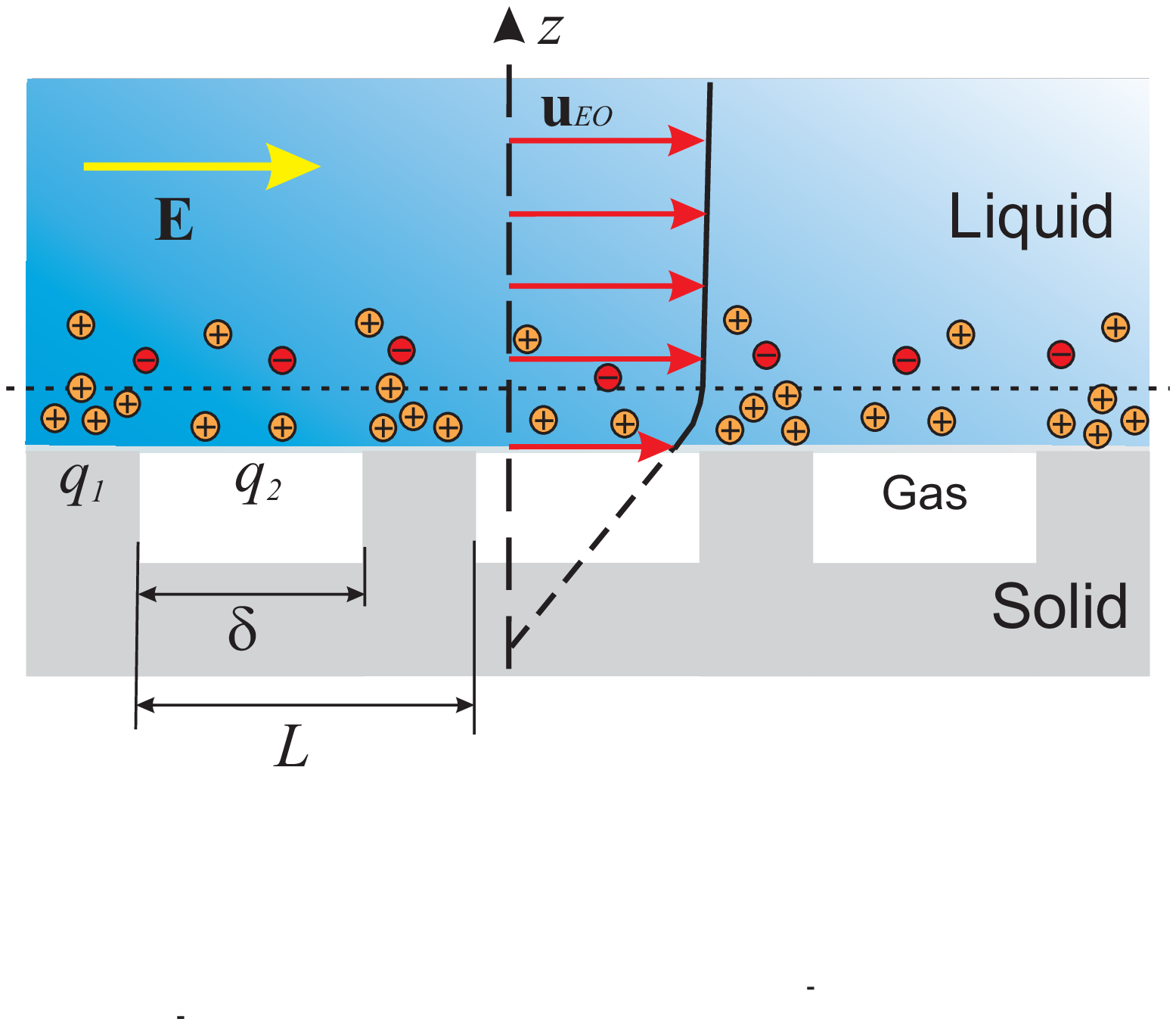}
  \end{center}
  \caption{Sketch of the influence of hydrophobic (a) and superhydrophobic (b) slippage on the electro-osmotic flow. }
  \label{fig:electrokinetics}
\end{figure}

Electro-osmosis (EO), i.e. flow generation by an
electric field, originally developed in colloid science, is currently experiencing a renaissance in
microfluidics. It may be considerably amplified by surface slippage, even for nanometric slip length. The reason
for this amplification is that the electric double layer (EDL), characterized by the Debye screening length
$\lambda_D=\kappa^{-1}$ defines an additional length scale of the problem comparable to $b$. According to classical formula~\cite{joly2004,muller.vm:1986}
\begin{equation}
u = -\frac{q_0 E_0}{\eta \kappa}\left(1 +  b \kappa \right) = -\frac{\epsilon \zeta  E_0}{\eta }\left(1 +  b \kappa \right)
\label{muller}
\end{equation}
where $u$ is the electro-osmotic velocity (outside of the double layer), $ q_0$ is the surface charge density, $\zeta =q_0/\kappa \epsilon$ is the zeta-potential across the diffuse (flowing) part of the double layer, $\epsilon$ is the permittivity of the solution, and $E_0$ is the tangential electric field. Therefore, the flow can potentially be enhanced for a thin compared to $b$ EDL, i.e. when $\kappa b \gg 1$ (see Fig.~\ref{fig:electrokinetics}), which has been proven experimentally~\cite{muller.vm:1986,bouzigues.c:2008}.

This expression can be generalized to other interfacial transport phenomena, such as diffusio-osmosis and thermo-osmosis, as it has been well discussed in~\cite{ajdari.a:2006,bocquet2007}. These correspond to the induction of a flow by the gradient of a solute
concentration for the former and by a gradient of temperature
for the latter~\cite{anderson.jl:1989}. The flow velocity for these two important phenomena is proportional to the applied gradient of concentration or temperature. Similarly to electro-osmosis, for both, slippage amplifies the velocity with the factor $(1+b/\lambda)$, where $\lambda$ is a thickness of a thin interface layer (of the order of the range of interaction of the solute with the solid surface).

\subsection{Super-hydrophobic surface}

It is now attractive to consider electro-osmotic flow over super-hydrophobic surfaces, whose texture can significantly amplify hydrodynamic slip. Eq.(\ref{muller}) suggests that a massive amplification of EO flow can be potentially achieved over super-hydrophobic surfaces. However, the controlled generation of such flows is by no means obvious since both the slip length and a charge distribution are inhomogeneous and anisotropic. Despite its fundamental and practical significance, electro-osmotic flow over super-hydrophobic surface has so far received little attention. Only recently such a flow has been investigated past inhomogeneously charged slipping surface in the case of a thick channel ($H \gg L$) and perfect slip ($b\to \infty$) at the gas sectors.~\cite{Squires08,bahga:2009}

\begin{figure}
\begin{center}
 \includegraphics [width=8 cm] {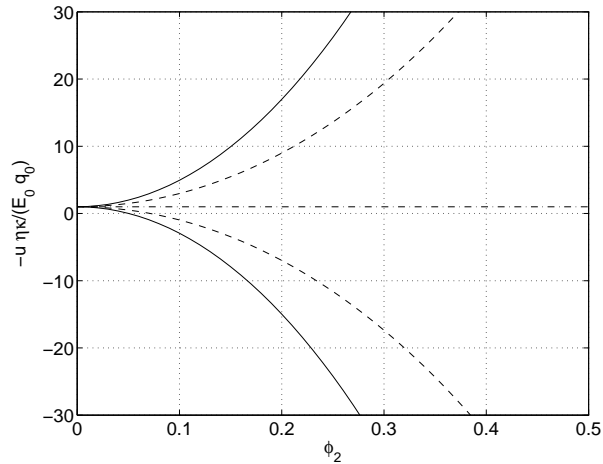}
   \end{center}
  \caption{The electro-osmotic slip velocities (thin EDL) for eigendirections of super-hydrophobic stripes. Dash-dotted line corresponds to the case of uncharged gas sectors $q_{1}=q_0$, $q_2=0$: transverse and longitudinal EO flows coincide with each other. Upper solid curve corresponds to longitudinal, and upper dashed - to transverse EO flow for a surface of a constant charge $q_{1}=q_2=q_0$.  Lower solid curve corresponds to longitudinal, and upper dashed - to transverse EO flow for a surface with $q_{1}=-q_2=-q_0$.}
  \label{fig:electrokin}
\end{figure}

The general result for a thin  EDL ($ \kappa L \gg 1$) has been formulated in~\cite{Squires08}
\begin{equation}
{\ub}  = -\frac{ \mathbf{E}_0}{\eta \kappa}\cdot\left(q_1 {\bf I} +  q_2 {\bf b}_{\rm eff} \kappa \right)
\label{squires}
\end{equation}
by using the Lorentz
reciprocal theorem for the Stokes flow. Here ${\bf I}$ is unity tensor, $q_1$ and $q_2$ are the surface charge density at the solid and gas regions, correspondingly.  This expression shows that surface anisotropy generally leads to a tensorial electro-osmotic mobility. To highlight the effect of anisotropy, we focus now on patterned super-hydrophobic surface consisting of periodic stripes, where the surface charge density varies only in one direction, likewise the slip lengths. By combining Eq.(\ref{squires}) with Eqs.(\ref{beff_ort_largeH_id}) we get general results and important limiting situations for stripes summarized in Table~\ref{tbl:EO}, and discussed below.

\begin{table}
  \small
  \tbl{Electro-osmotic slip past super-hydrophobic stripes in a thin EDL limit}
  {\begin{tabular}{l|cc}
  \hline
         &  $u_{\rm eff}^\parallel$  &  $u_{\rm eff}^\perp$    \\
  \hline
   \emph{General case}& $-\displaystyle\frac{E_0}{\eta\kappa}\left(q_{1} +  q_2 \displaystyle\frac{\kappa L}{\pi}\ln\left[\sec\left(\displaystyle\frac{\pi \phi_2}{2}\right)\right] \right)$ &  $-\displaystyle\frac{E_0}{\eta\kappa}\left(q_{1} +  q_2 \displaystyle\frac{\kappa L}{2\pi}\ln\left[\sec\left(\displaystyle\frac{\pi \phi_2}{2}\right)\right] \right)$  \\
  \emph{$q_1 = q_0,\,\,q_2=0$}& $-\displaystyle\frac{E_0 q_{0}}{\eta\kappa}$ &  $-\displaystyle\frac{E_0 q_{0}}{\eta\kappa}$ \\
  \emph{$q_1 = q_2= q_0$}  &  $-\displaystyle\frac{E_0 q_0}{\eta\kappa}\left(1 +  \displaystyle\frac{\kappa L}{\pi}\ln\left[\sec\left(\displaystyle\frac{\pi \phi_2}{2}\right)\right] \right)$ &  $-\displaystyle\frac{E_0 q_0}{\eta\kappa}\left(1 +   \displaystyle\frac{\kappa L}{2\pi}\ln\left[\sec\left(\displaystyle\frac{\pi \phi_2}{2}\right)\right] \right)$   \\
  \emph{$q_1 = - q_2= q_0$}   &  $-\displaystyle\frac{E_0 q_0}{\eta\kappa}\left(1 -  \displaystyle\frac{\kappa L}{\pi}\ln\left[\sec\left(\displaystyle\frac{\pi \phi_2}{2}\right)\right] \right)$  &   $-\displaystyle\frac{E_0 q_0}{\eta\kappa}\left(1 -   \displaystyle\frac{\kappa L}{2\pi}\ln\left[\sec\left(\displaystyle\frac{\pi \phi_2}{2}\right)\right] \right)$ \\
  \hline

  \end{tabular}}  \label{tbl:EO}
\end{table}

   In case of uncharged gas interface ($q_{1}=q_0$, $q_2=0$) we predict the simple Smoluchowski formula~\cite{Squires08}. In other words, there is no EO flow enhancement, and the flow is isotropic despite anisotropy of a surface (see Fig.~\ref{fig:electrokin}). This surprising result has been confirmed by molecular dynamic simulations~\cite{Huang08} and later analysis~\cite{bahga:2009}. Note that for uncharged gas interface with thick EDL ($ \kappa L \ll 1$), the results are qualitatively different, and for uncharged gas interface the EO flow remains tensorial~\cite{bahga:2009}. However, since $ \kappa L$ is small and since the electro-osmotic velocity is proportional to the fraction of a charged area,$\phi_1$,  electro-osmotic flow becomes suppressed (compared to predicted by the Smoluchowski equation) despite a large effective slip~\cite{vinogradova.oi:2010}.

   In case of a charged gas interface (only) a considerable enhancement of electro-osmotic flow is possible. For a uniformly charged ($q_{1}=q_2=q_0$) anisotropic super-hydrophobic surface the expression for electro-osmotic flow can be transformed to
  \begin{equation}
{\ub} = -\frac{ \mathbf{E}_0 q_0}{\eta \kappa}\cdot\left({\bf I} +  {\bf b}_{\rm eff} \kappa \right),
\label{squires2}
\end{equation}
   which might be seen as a tensorial analog of Eq.~(\ref{muller}). Fig.~\ref{fig:electrokin} includes theoretical results calculated with Eq.~(\ref{squires2}) for a geometry of stripes, and is intended to demonstrate that the flow is truly anisotropic and can exhibit a large enhancement from effective hydrodynamic slip, possibly by an order of magnitudes. We stress that such an enhancement is possible even at a relatively low gas fraction, i.e. when ${\bf b}_{\rm eff}$ is relatively small (but the amplification ratio, $ {\bf b}_{\rm eff} / \lambda_D$, might be huge). Obviously, in case of isotropic super-hydrophobic surface, Eq.~(\ref{squires2}) transforms to
 \begin{equation}
u = -\frac{ E_0 q_0}{\eta \kappa}\left(1 +  {b}_{\rm eff} \kappa \right),
\label{squires3}
\end{equation}
and the amplification of EO flow at the isotropic super-hydrophobic surface (such as observed in a recent experiment~\cite{audry.mc:2010}) might serve as a very strong evidence in favor of a charge at the liquid-gas interface.

An interesting scenario is expected for oppositely charged solid and gas sectors. In this case Eq.~(\ref{squires}) transforms to
\begin{equation}
{\ub} = -\frac{ \mathbf{E}_0 q_0}{\eta \kappa}\cdot\left({\bf I} -  {\bf b}_{\rm eff} \kappa \right).
\label{squires4}
\end{equation}
The calculation results for this situation are also included into Fig.~\ref{fig:electrokin}, and suggest a very rich fluid behavior. We see, in particular, that inhomogeneous  surface charge can induce EO flow along and opposite to the field, depending on the fraction of the slipping area. Already a very small fraction of the gas sectors would be enough to reverse the effective EO flow. Another striking result is that electro-neutral surface ($\langle q\rangle = \phi_1 q_1 +\phi_2 q_2 = 0$) can generate extremely large EO slip. With our numerical example this corresponds to $\phi_2=0.5$.

Of course, the area of research connected with interface transport phenomena is still at its infancy. Thus, the role of the surface conductance just started to be probed~\cite{zhao.h:2010}. Beside that, many assumptions exploited above should obviously be relaxed. For example, in future, we suggest as a fruitful direction to consider electro-osmotic flow in a thin gap and by assuming a partial slip at the gas sectors.  It will be very important to investigate transverse electrokinetic phenomena, that could be greatly amplified by using striped super-hydrophobic surfaces. Note, that if the charge is varied along the direction of electric field, the fluid close to a super-hydrophobic wall is pulled periodically in opposite directions. As a result, the recirculation rolls should develop on a scale proportional to a texture size, $L$. This should provide an additional opportunity for a mixing, similar to described in~\cite{ajdari.a:1995}, but hopefully much faster.

\section{Conclusion and future directions}

With recent progress in micro- and nanofluidics new interest has arisen in determining forms of hydrodynamic boundary conditions~\cite{stone2004,Bazant08,kamrin.k:2010}. In particular, advances in lithography to pattern substrates have raised several questions in the modeling of the liquid motions over these surfaces and led to the concept of the effective tensorial slip. These effective conditions capture complicated effects of surface anisotropy and can be used to quantify the flow over complex textures without the tedium of enforcing real inhomogeneous boundary conditions.

This chapter has discussed the issue of boundary conditions at smooth hydrophobic and rough hydrophilic surfaces, and has then given the especial emphasis to the effective boundary conditions for a flow past hydrophobic solid surfaces with special textures that can exhibit greatly enhanced (`super') properties, compared to analogous flat or slightly disordered surfaces. Research on these super-hydrophobic materials during past decades  has mostly focused on their extreme non-wettability~\cite{quere.d:2005,quere.d:2008}. However, now the field has moved beyond wetting towards transport phenomena~\cite{vinogradova.oi:2010,bocquet2011}. An effective slip becomes the main parameter to quantify the effective transport properties near super-hydrophobic surfaces, which is in contrast to a  traditional approach, based on the use of the effective contact angle.

We have discussed formulas describing effective boundary conditions for pressure-driven flow past super-hydrophobic textures of special interest (such as stripes, fractal patterns of nested circles, chessboards, and more). The predicted large effective slip of super-hydrophobic surfaces compared to simple, smooth channels can greatly lower the viscous drag of thin microchannels.

The power of the tensor formalism and the concept of effective slippage has then been  demonstrated by exact solutions for two other potential applications: optimization of the transverse flow and analytical results for the hydrodynamic resistance to approach of two surfaces. These examples demonstrate that properly designed super-hydrophobic surfaces could generate a very strong transverse flow and significantly reduce the so-called `viscous adhesion'. Finally, we have discussed how super-hydrophobic surfaces could amplify electrokinetic pumping in microfluidic devices.

As we have shown, a combination of wetting and roughness provides many new and very special hydrodynamic properties of surfaces, which could be explored with the formalism discussed here. This should allow the local slip tensors to be determined by global measurements, such
as the permeability of a textured channel as a function of the surface orientations or the hydrodynamic force exserted on the body approaching a super-hydrophobic plates. They may also guide the design of super-hydrophobic surfaces for microfluidic lab-on-a-chip and other applications.

Despite the impressive advances in hydrodynamic slippage phenomena, many
challenges remain, both theoretical and experimental, fundamental
and practical. In particular, there is quantitative discrepancy between theory and experiment (measured effective slip are persistently lower than the theoretical predictions). This discrepancy reflects gaps in our fundamental understanding of flow past rough and hydrophobic surfaces. Below we briefly discuss promising future directions for research,
both experimental and theoretical.

\p    Experiments in thick channels~\cite{tsai.p:2009,ou2005} have established that hydrodynamic flows are generally slower than one would expect from theory.~\cite{vinogradova.oi:2010}  Current analytical models of super-hydrophobic effective slip are based on the idealized model of a heterogeneous surface with patches of boundary conditions, and mostly neglect a number of dissipation mechanisms in the gas phase and at the interface. The effects associated with different aspects of the gas flow and meniscus curvature must be included in the models. Regardless recent semi-analytical and numerical analysis~\cite{ng.co:2010,sbragaglia.m:2007} the goal should remain to find simple analytical formulas, with as few adjustable parameters as possible, to fit experimental data.

\p Drainage experiments conducted in the AFM and SFA,
which has the ability to probe fluid films of nano- and molecular thicknesses,
may be able to yield detailed information on super-hydrophobic slip. Some data are already available,~\cite{wang.y:2010,wang.y:2009,steinberger.a:2007} but we are unaware of any previous
work that has studied how the squeeze film drainage between
curved surfaces would be modified by the occurrence of the effective slip.
Recent theory of hydrodynamic interaction between disks~\cite{belyaev.av:2010b} shed some light on what could happen qualitatively, but cannot be used for a quantitative analysis of the hydrodynamic data obtained with the AFM and SFA. The same remark concerns the use of a theory of a film drainage between smooth hydrophobic surfaces~\cite{vinogradova.oi:1995a}, which is not fully applicable to quantify a super-hydrophobic slip. We believe that a challenge for a theory would be to develop a theoretical modeling of experimentally relevant sphere vs. plane geometry. This will open many possibilities for new experiments, and could revolutionize the field.

\p There are many opportunities to design new experiments and to develop improved theoretical models for electro-osmotic flow past rough, hydrophobic and super-hydrophobic surfaces.~\cite{bahga:2009,Squires08} For example, a systematic study of the effect of surface texture, amplitudes of local slip lengths, and a role of a channel thickness would be interesting. It would likewise be interesting to work with conducting, but hydrophobic or super-hydrophobic surfaces. The study of super-hydrophobic Cassie surfaces will naturally pose fundamental questions about a mechanism of electro-osmosis at the charged gas interface, where both adsorbed ions and their screening clouds are mobile.~\cite{bahga:2009}

\p  Very promising directions are certainly the diffusio-osmosis and thermo-osmosis, where solvent flows are induced by gradients of solute concentration or temperature. Being combined with hydrophobic~\cite{ajdari.a:2006} or super-hydrophobic slippage, these effects could lead to a giant amplification of flows in microchannels, even if the liquid-gas interface is uncharged as shown in recent simulation study.~\cite{Huang08} The combination of these two strategies, i.e. diffusio-/thermo-osmosis and super-hydrophobicity, has to be studied theoretically and experimentally, and we expect a significant expansion into this very interesting area of research.


\bibliographystyle{psp-rv-van}
\bibliography{slip}

\printindex                         
\end{document}